\documentclass[a4paper, 11pt]{article} 
\usepackage{amsfonts,amssymb,amsmath}
\usepackage{epsfig,epsf}
\usepackage{graphicx}
\usepackage{subcaption}
\usepackage{a4wide}
\usepackage{amsmath}
\usepackage{mathtools}
\usepackage{bbm}
\usepackage{amsthm}
\usepackage[toc,page]{appendix}
\usepackage{afterpage}
\usepackage{enumitem}

\usepackage{xcolor}
\usepackage{url}

\usepackage{authblk}
\usepackage{orcidlink}
\usepackage{cleveref}

\allowdisplaybreaks

\usepackage[round]{natbib} 
\bibpunct{(}{)}{;}{a}{}{,} 

\setlist[enumerate,1]{label=\arabic*.,ref=\arabic*}
\setlist[enumerate,2]{label=\alph*.,ref=\arabic{enumi}.\alph*}
\setlist[enumerate,3]{label=\roman*.,ref=\arabic{enumi}.\alph{enumii}.\roman*}

\numberwithin{equation}{section}

\allowdisplaybreaks

\numberwithin{figure}{section}
\newtheorem{thm}{Theorem}[section]

\theoremstyle{definition}
\newtheorem{defn}[thm]{Definition}

\newcommand{\Real}{\mathbb R}

\newcommand{\diag}{\mathrm{diag}}
\newcommand{\argmax}{\mathrm{argmax}}

\newcommand*{\ud}{\mathrm{d}}
\newcommand{\pspace}{(\Omega,\mathcal{F},\mathbb{P})}

\newcommand{\dd}{\mathrm{d}}

\newcommand{\pd}{\mathrm{PD}}

\newtheorem{theorem}{Theorem}[section]

\theoremstyle{definition}

\newtheorem{algorithm}[theorem]{Algorithm}

\begin{document}

\title{Calibration of the rating transition model for high and low default portfolios}



\def\correspondingauthor{\footnote{Corresponding author: j.he2@uva.nl}}

\author[1]{Jian He \correspondingauthor{}}
\affil[1]{Korteweg-de Vries Institute, University of Amsterdam, Amsterdam, The Netherlands}
\author[1]{Asma Khedher }
\author[1, 2]{Peter Spreij}
\affil[2]{IMAPP, Radboud University, Nijmegen, The Netherlands}

\maketitle

\begin{abstract}
\noindent
In this paper we develop Maximum likelihood (ML) based algorithms to calibrate the model parameters in credit rating transition models. Since the credit rating transition models are not Gaussian linear models, the celebrated Kalman filter is not suitable to compute the likelihood of observed migrations. Therefore, we develop a Laplace approximation of the likelihood function and as a result the Kalman filter can be used in the end to compute the likelihood function. This approach is applied to so-called high-default portfolios, in which the number of migrations (defaults) is large enough to obtain high accuracy of the Laplace approximation. By contrast,  low-default portfolios have a limited number of observed migrations (defaults). Therefore, in order to calibrate  low-default portfolios, we develop a ML algorithm using a particle filter (PF) and Gaussian process regression. Experiments show that both algorithms are efficient and produce accurate approximations of the likelihood function and the ML estimates of the model parameters.
\smallskip\\
{\sl keywords: credit risk, transition model, parameter calibration, Kalman filter, particle filter, Laplace method} 
\end{abstract}

\newpage
\tableofcontents
\newpage

\section{Introduction}

\subsection{Problem description and background}

Credit transitions, also referred to as credit migrations, constitute one of the most important building blocks of credit risk management. It concerns the change of the creditworthiness of a firm or a particular debt issue, which leads to potential credit losses and hence result in credit risk. The creditworthiness is usually measured by credit ratings (or scores). For instance, Standard \& Poor’s Rating Services, Moody’s Investor Services, and Fitch Ratings assign corporate bond issuers with credit ratings to reflect their creditworthiness. 
Once the ratings are defined, the objective is to determine the probability with which the credit risk rating of a borrower decreases or increases by a certain degree, from one level to another, lower or higher, one. The  probabilities, that a credit risk rating of a borrower decrease or increase from one period to the next one, are usually collected in a \emph{transition matrix}. Namely, the transition matrix lists the probabilities of the migrations between different credit ratings over specific time intervals. A temporal analysis of credit migrations requires a dynamic model of the transition matrix, the transition model. 
One of the focuses of the transition model is to address the \emph{cross-sectional dependence} in default rates within a time period due to common economic conditions, i.e.\ systematic risk factors. This cross-sectional dependence especially describes how likely the obligors default together in a stress scenario and is hence essential in credit risk measuring. The classic examples of modelling the cross-sectional dependence are the Merton model \citet{merton1974pricing} and its extensions such as \citet{CreditMetrics1997} and \citet{Moody's2003} from Moody's. 
These models belong to the class of so-called structural models, in which the systematic risk factors together with the obligor specific idiosyncratic factors reflect the obligor's asset value. 

A default occurs when the asset value falls below a certain threshold, the default barrier. In these models, the sensitivity of the asset value to the systematic risk factors determines the asset value correlations, and hence also default correlations between different obligors. This asset value correlation is usually referred to as the \emph{asset correlation}.  
Structural models can also be generalized to describe rating migrations. This requires a definition of rating barriers, and rating changes occur when the asset value crosses a rating barrier value. The dynamics of a systematic risk factor are specified by a model. The model can be a latent variable model, such as the Merton model \citet{merton1974pricing},  \citet{albanese2003credit}, or a mixture model which uses both latent and observed (Macroeconomic) factors to form the systematic risk factors, see \citet{CreditRisk+}, \citet{mcneil2007bayesian}.  

The risk factor model parameters (if there is any), the asset correlation, and the rating barriers can be calibrated to time series of historical migration or market data. Due to the presence of the latent variables, the likelihood of the joint default and migration events depends on an integral, which in general lacks a closed form formula. Therefore, in practice, the latent variables at different periods are usually assumed to be independent to simplify the integral in the likelihood function, so that numerical integration and maximum likelihood (ML) methods can be relatively easily  be implemented, see for instance \citet{gordy2002estimating}, \citet{frey2002dependence}, \citet{demey2007maximum}. 
On the other hand, since it is well observed that the default (migration) intensity varies over time according to an economic cycle (see, for instance, \citet{mcneil2007bayesian}), a good risk factor model is required to capture \emph{serial dependence} caused by the cyclical behaviour of economy. However, including  serial dependence significantly increases the complexity of the integral in the likelihood function, and consequently makes  ML estimation very difficult to perform by using numerical integration. 

There are few attempts in the literature to fit such a model with serial dependent latent processes, which belongs to the category of \emph{state space models}, to default or migration data. \citet{koopman2005} consider a model where default event is driven by continuous latent factors. But they model the default rates (i.e.\ the ratio of default obligors) instead of the default counts. This simplification leads to a much less complex likelihood function, but it requires more model assumptions and may show undesirable features, especially when the number of obligors or the number of defaults is small. In order to calibrate the transition model directly to default counts rather than to default rates, \citet{mcneil2007bayesian} implement a Markov chain Monte Carlo (MCMC) approach, more specifically Gibbs sampling, to calibrate the mixture transition model. Although the MCMC approach has its advantages, in general it is slow. It requires tuning the burn in period and simulating nested Monte Carlo samples, i.e.\ first simulating the model parameters and then simulating the latent states given each simulated model parameter. The running time of an MCMC algorithm is acceptable when only calibrating a few transition models. However, when dealing with large portfolios which contain a lot of different rating systems, the MCMC approach is impractical.

\subsection{Contributions}

The main contributions of this paper are two new calibration algorithms designed for high and low default portfolios respectively. A higher default portfolio, usually a retail portfolio, is a portfolio with a relatively large number of observed defaults, which means that either the probability of default is relatively high or the number of obligors (observations) is large. By contrast, a low default portfolio, such as the portfolio of large corporate or financial institutions, means a portfolio with few observed defaults, due to a small default probability and a small amount of obligors in the portfolio. 

The two new calibration methodologies aim to calibrate the transition model to the observed default or migration counts. For a high default portfolio, we propose a calibration approach which approximates the likelihood function by using the Laplace method. We show that, as a consequence, the Kalman filter can be used to compute the approximated likelihood functions. We also show that mode estimation for the Laplace approximation and the log-likelihood function can be calculated by the same Kalman filter iterations. Therefore numerical ML estimation is extremely fast based on the proposed Laplace approach. 
For the low default portfolio, due to the small number of observations, the Laplace approximation is less accurate and hence can introduce bias in the parameter estimates. In this case, we propose a Monte Carlo approach, more precisely, a particle filter, to estimate the likelihood function. This proposed Monte Carlo approach leverages on the mode estimation algorithm used in the Laplace approach, and uses its outcome to construct the importance density for the Monte Carlo sampling. The Monte Carlo approach is very efficient and provides very accurate estimation of the likelihood function, with small amounts of Monte Carlo samples. Moreover, since the likelihood function estimated by the Monte Carlo approach is not continuous w.r.t.\ the model parameter, we also implement a Gaussian process regression, a grid based approach, to smooth the likelihood function so that the ML algorithm can be easily applied. Since the proposed algorithm does not require nested Monte Carlo simulations, it is much faster than a conventional MCMC approach.

The proposed calibration methodologies are generic. They can not only deal with default models, which most literature usually focuses on, but also with migration models. They can also be applied to various structural or mixture transition models with different response functions, such as probit or logistic functions. For instance, it can be applied to calibrate the transition model in \citet{He2023}. Numerical experiments are conducted in this paper to assess the performance of the proposed calibration methodologies. The experiments use simulated migration data, based on the model parameters with certain specified values. These parameters with specified values are considered to be the `true' values and used as a benchmark when comparing with the values from the calibration. The results indicate that both proposed calibration methodologies are efficient and provide accurate approximations on the likelihood function and estimates of the model parameters.   

\subsection{Organization of the paper}

Section~\ref{section:set-up} provides the background knowledge of the migration models in credit risk modelling and presents the migration models studied in this paper. The proposed ML methodologies are presented in Section~\ref{section:calibration}. In Section~\ref{section:numerics} we provide numerical experiments to show the performance of the proposed approaches. Finally, Section~\ref{section:conclusions} is devoted to the conclusions. In the Appendix we summarize some results on the state space model and on Bayesian filters.

\subsection{Notation and conventions}\label{sec:notation-and-conventions}

Here follows some notations and other conventions used in this paper. All random processes are defined on a fixed probability space $(\Omega, \mathbb{F}, \mathbb{P})$. We will use discrete time, a typical period denoted by $k$. The discrete time filtration is $\{\mathcal{F}_k, k=0, 1, \ldots, n\}$, where $\mathcal{F}_k$ summarizes the information up to time $k$ and $n$ is the final time. All processes are assumed to be adapted to this filtration. We denote by $\mathbb{N}^+$ the strictly positive integers. By $\cdot^\top$ we denote the transpose of a vector or a matrix. 

We consider a migration matrix with $R \in \mathbb{N}$ different ratings, including default as the last rating. We denote by $T_{ij,k}$ the probability of the transition from a credit state $i=1,\ldots, R$ at time $k$ to a credit state $j=1,\ldots,R$ at time $k+1$ . Similarly, the number of (observed) migrations from rating $i$ at period $k$ to rating $j$ at period $k+1$ is denoted as $m_{ij, k}$. Therefore, the migration matrix at period $k$, denoted by $M_k$, is defined as $M_k=(m_{ij, k}, i=1, \ldots, R, j=1, \ldots, R) = (m_{1, k}^\top, \ldots, m_{R, k}^\top)^\top$, where $m_{i, k}=m_{i, k}=(m_{i1, k}, \ldots, m_{iR, k})$ denotes row $i$ of the migration matrix $M_k$. We denote by $M=(M_1, \ldots, M_n)$ the (observed) sequence of migration matrices of $n$ periods. We use similar notations for other matrices. We denote by $x_{1:n}$ a sequence $x_1,\ldots, x_n$. 

Finally, we collect all  unknown model parameters in a vector $\psi$.

\section{Transition model}\label{section:set-up}

\subsection{Transition model review}

The probabilities that a credit risk rating of a borrower decrease or increase by a certain degree, from one period to the next one, are described by a transition matrix. Namely, the transition matrix presents the probabilities of the migrations between different credit ratings over specific time intervals. 
In the transition matrix, the element located at the $i$-th row and $j$-th column represents the probability of a borrower migrates from the $i$-th rating to the $j$-th rating. These are called transition probabilities and are often related to macroeconomic variables such as interest rates, inflation, gross domestic product (GDP), unemployment, etc. Alternatively, transition probabilities can also be modelled by using certain abstract latent processes. 

The transition models can in general be divided into two main classes, structural and reduced form models. Structural models link the up- or down-grade (or default) probabilities of a firm to the value of its assets and liabilities.  The most popular link functions are probits and logistic functions. The structural models were pioneered by \citet{merton1974pricing}, usually referred to as the Merton model, which applies principles of Black and Scholes option pricing in corporate debt valuation. This approach models equity as a call option on the underlying assets of a firm with strike value equal to the book value of the firm's liabilities. In particular, a default event is triggered if the firm's asset value drops below its liabilities. In this case, the correlations between different firms in a default event is a result of the correlated asset values. Therefore, the so-called asset correlation is usually used to describe the correlation in the default event. The Merton model is further developed by, for instance, \citet{hull2001valuing}, \citet{avellaneda2001distance}, \citet{duffie2004credit} and \citet{albanese2003credit}. 

In these structural models, a latent credit process $X_t$, called a credit index, is used to reflect unobservable credit quality over time which is driven by firm-specific variables such as the asset values. A default occurs if the credit index $X_t$ crosses a default barrier, which is often interpreteted as the value of the firms liabilities. The dynamics of the credit index are specified by the model and the default barriers can be calibrated to historical migration or market data. Structural models can be generalized to describe not only defaults, but also rating migrations. This requires a mapping from the latent credit index to the rating states at each time, which corresponds to specifying an interval for the credit index that corresponds to each rating class. The intervals are defined by rating barriers, and rating changes occur when the credit index passes a barrier value. 

Reduced form models, by contrast, treat a default as an unexpected event whose probability is governed by a default intensity process. Reduced form models originated with \citet{jarrow1992credit}, and later studied by, for example, \citet{jarrow1995pricing}, \citet{duffie2004credit}, \citet{frey2002dependence}, \citet{wendin2006dependent} and \citet{koopman2008multi}. Unlike the structural model, in which the default relies on a firm's capital structure, the reduced form model describes the default as an event of a jump process driven by exogenous factors. In such models, default correlation is captured by co-movement in default intensities. The reduced form model may be easy to be implemented and calibrated as no estimation of a firm's asset or liabilities value is required. However, these models suffer from the lack of economic interpretation about default behaviour. Whilst, structural models are particularly useful in credit risk management field because they provide intuitive economic interpretations of default events. In fact, this approach, for example, is used in the Vasicek formula, see \citet{Vasicek2002}, an example of an asymptotic single factor model, which is the cornerstone of the IRB (internal rating-based approach) formula of a regulatory credit risk risk weighted assets (RWA) calculation. A structural approach also facilitates the calibration of the model using various data sources, such as default rates, rating transitions, equity returns (as a proxy for asset returns) and CDS spreads. The latter is commonly applied when calibrating sovereign portfolios where the numbers of observations are extremely limited.  

The key role of the transition model is to describe the dependence of the migrations (defaults). One important dependence is the cross-sectional dependence, which models the correlations of the credit rating movements among or within different (homogeneous groups of) obligors (e.g., obligors could be grouped by country, industry, rating, etc.). These dependencies are typically caused by exposure to common systemic risk factors. Articles such as \citet{nickell2000stability} and \citet{bangia2002ratings} demonstrated that besides cross-sectional correlations, default (migration) rates also exhibit serial dependence. Serial correlation is a result of cyclical behaviour of economic factors. In particular, it implies that a poor (good) economic state is more likely to be followed by a poor (good) economic year and decreased (increased) asset values, and in the end increase (decrease) the change of default or downward migrations.  Factors used to explain variation in migration rates can be classified into observed (macroeconomic) and unobserved (latent) factors. The use of macroeconomic factors is motivated by the observation that default rates in the financial, corporate, and household sectors increase during recessions. This observation leads to the implementation of credit models that attempt to explain default indicators using economic variables. For example, \citet{simons2018macroeconomic} uses GDP, interest rates, exchange rates and oil price to explain default rates. An advantage of a macroeconomic factor model is that this type of model is very suitable for designing stress scenarios. However, macroeconomic variables are business cycle indicators and may not be an optimal proxy for a credit cycle, see for example \citet{gorton2008bank} and \citet{koopman2005business}. Furthermore, macroeconomic variables have to be modelled if used for simulation of migration rates. An alternative approach is to employ unobserved factors. In this case the dynamics of any underlying systematic component are estimated directly from the data. Literature which allow for unobserved factors includes for instance \citet{koopman2008multi}, \citet{gagliardini2005stochastic} and \citet{mcneil2007bayesian}. While latent factor models are less prone to misspecification of a credit cycle, more advanced mathematical methods, such as the Kalman filter or a particle filter, are required to estimate them. 

\subsection{Specified transition model}\label{subsection:transition-probabilities}

Structural models are usually used in measuring the credit risk since such models have an economic interpretation, and parts of the models can be calibrated to external data. A mixture model for transitions including both observed and latent systematic factors is introduced in this paper. This type of the model is an example of a \emph{general linear mixture model} (GLMM), see  \citet{mcneil2007bayesian}. The precise specification is as follows.

Suppose we have $R\geq 2$ credit states, in which the last state $R$ is default. Recall $T_{ij,k}$, $i,j=1,\ldots, R$ and $k=1,\ldots,n$ as defined in Section~\ref{sec:notation-and-conventions}.  We model the transition probability $T_{ij,k}$, $i,j=1\ldots, R$ at period $k$ as \footnote{In some cases, the cumulative transition probability, instead of the transition probability, is modelled through the response function. One of the examples can be found in Section~\ref{subsubsection:example2}.}
\begin{equation}\label{eq:tg}
T_{ij,k} = g(\theta_{ij, k}).
\end{equation}
Here 
$g: \mathbb{R}\rightarrow[0, 1]$ is a so-called response function. The $\theta_{ij,k}$, $i,j=1, \ldots, n$ are often referred to as the signals that describe the dynamics of the transition probabilities at period $k$. We assume the signals are linearly driven by the {\it latent} common factors $(x_k)_{k=1,\ldots, n}$ and the {\it observed} common factors $(u_k)_{k=1,\ldots, n}$ as 
\begin{equation}\label{eq:migration}
\theta_{ij, k} = d_{i,j} + K_{i,j}^\top x_k + L_{i,j}^\top u_k, \qquad i, j=1, \ldots, R\,.
\end{equation}
Here the constants $d_{i, j}$ indicate the level of the (cumulative) transition probabilities from rating $i$ to rating $j$, the latent common factors $x_k$ at period $k$ are $s\times 1$ random vectors, the observed common factors $u_k$ at period $k$ are $l\times 1$ vectors which present certain predetermined macro-economic variables or market indices that can be directly observed in the market. The $K_{i,j}$ and $L_{i,j}$, for every $i,j=1, \ldots, R$ are the factor loadings with dimension $s\times 1$ and $l \times 1$ respectively. They describes the sensitivity of the signal to the common factors. The latent process $x$ is assumed to be an autoregressive  AR(1) process (see, e.g.~\citet[Chapter 3]{BrockwellDavis} for an introduction to such models) which evolves linearly over time with a Gaussian error $\eta$.
Note that the migrations $\theta_{ij,k}$, $i,j=1,\ldots,R$ at period $k$ as defined in \eqref{eq:migration} are correlated.  

Recall $m_{i,k}$, $i=1,\ldots, R$, $k =1, \ldots,n$ as introduced in Section~\ref{sec:notation-and-conventions}. It is obvious that the random vector $m_{i, k}$ with dimension $R\times 1$ follows a multinomial distribution when $T_{ij, k}$ is given. Since $T_{ij, k}$ is assumed to be linked with the signal $\theta_{ij, k}$ through the response function $g$, our transition model in the end is given by for $k=1,\ldots,n\,$, $i,j=1,\ldots,R$\,,
\begin{equation}\label{eq:model-m-theta}
\begin{aligned}
m_{i, k} &\sim mn\left(\cdot \mid N_{i, k}, g(\theta_{ij, k})\right), \\
\quad  \theta_{ij, k}  &= d_{ij} + K_{i, j}^\top x_k + L_{i,j}^\top u_k, \\
x_{k} & = A x_{k-1} + \eta_{k}. 
\end{aligned}
\end{equation}
In \eqref{eq:model-m-theta} $mn$ stands for multinomial distribution, $N_{i, k}=\sum_{j=1}^R m_{ij, k}$ is the number of independent trials (i.e.\ the number of clients) in period $k$ with rating $i$. The initial distribution of $x_k$ at a period $k$ is assumed to be Gaussian with mean $a_0$ and covariance matrix $P_0$, i.e., $x_0 \sim \mathcal{N}(a_0,P_0)$. The $\eta_{k}$ are assumed to be independent with a common $\mathcal{N}(0,Q)$ distribution, $A$ and $Q$ are $s \times s$-matrices.
Define $d=(d_{i, j}, i, j=1,\ldots,R)$, $K=(K_{i, j}, i, j=1,\ldots,R)$ and $L=(L_{i, j}, i, j=1,\ldots,R)$. The parameters $d$, $K$, $L$, $A$ and $Q$ are referred to as the parameters of the transition model and collectively denoted $\psi$.  The random vectors $m_{1, k}, \ldots, m_{R, k}$ are assumed to be independent given the signals $\theta=(\theta_{ij, k}, i,j=1, \ldots, R)$.

The transition model~\eqref{eq:model-m-theta} above belongs to the class of state space models, see Section~\ref{appendix:ssm}. We are interested in estimating the (latent) state process $(x_k)_{k=1, \ldots, n}$, but only have access to the process $(M_k)_{k=1, \ldots, n}$ as introduced in Section~\ref{sec:notation-and-conventions} which represents the observations. Because of the existence of the white noise in the data, estimating the value of the latent states $(x_k)_{k=1, \ldots, n}$ by the observations $(M_k)_{k=1, \ldots, n}$ is not trivial. There are different methods available in the literature to estimate the latent process, see e.g.~\citet{Press}, \citet{ChuiChen}, \citet{arulampalam2002tutorial}. The Bayesian filters and its extensions and the particle filter are the mostly used approaches in practice, see  Sections~\ref{subsection:KF} and \ref{subsection:PF} for an introduction and references to Kalman and particle filters. In particular, the Bayesian filters are also widely used to compute, or estimate, the likelihood function of the observations. In this paper, the Kalman filter and the particle filter are used as the building blocks in our proposed MLE based methodologies to estimate the unknown parameters in the model.

\subsection{Likelihood function of the observed migrations}

Similar to the notation of the migration matrices, we denote all the signals of the $n$ periods by $\theta=\{\theta_1, \ldots, \theta_n\}$, in which $\theta_k = \{\theta_{ij, k}, i, j=1, \ldots, R\}$ is the signal at period $k$. Note that the likelihood of the observed migrations $p(M|\psi)$ with respect to the unknown model parameter $\psi$ is obtained by computing the integral
\begin{align}\label{eq:calibration-psi}
p(M|\psi) &= \int p(M|\psi,\theta)p(\theta|\psi)\,\dd\theta\nonumber\\
 &= \int p(M|\theta)p(\theta|\psi)\,\dd\theta\,,
\end{align}
where we used the generic symbol $p$ to denote probabilities and densities.
Since the distribution of the migrations are assumed to be independent given the signal $\theta$, one obtains
\begin{equation}\label{eq:M_theta}
\begin{aligned}
p(M|\theta) &= \prod_{k=1}^{n} p(M_{k}|\theta_{k}) \\
&= \prod_{k=1}^{n} \prod_{i=1}^{R}p(m_{i, k}|\theta_{i, k})\,. 
\end{aligned}
\end{equation}
Plugging Equation~\eqref{eq:M_theta} into Equation~\eqref{eq:calibration-psi}, one obtains the likelihood
\begin{equation}\label{eq:likelihood}
\begin{aligned}
p(M|\psi) 
&=\int \prod_{k=1}^{n} \prod_{i=1}^{R}p(m_{i, k}|\theta_{i, k})p(\theta|\psi)\,\dd\theta. 
\end{aligned}
\end{equation}
If the $m_{i, k}$ would be linear in $\theta_{i, k}$, for $i=1,\ldots, R$, $k=1,\ldots, n$ and follow a Gaussian distribution, then the integrals in Equation~\eqref{eq:likelihood} have an analytic expression and can be calculated using the Kalman filter algorithm. However, the migrations $m_{i, k}$, for $i=1,\ldots, R$, $k=1,\ldots, n$ follow a multinomial distribution given the transition probabilities $T_{ij, k}, j=1, \ldots, R$. Therefore, there is no analytical expression for the integrals in \eqref{eq:likelihood}. Approximations or a Monte Carlo based approach are needed to estimate the likelihood function. In Section~\ref{section:calibration}, we propose two approaches, a Laplace approximation approach and a Monte Carlo based approach, to approximate the likelihood function. In the Laplace approximation approach, the Laplace method is applied to Equation~\ref{eq:calibration-psi} so that a Kalman filter can be applied, whereas the Monte Carlo approach uses the particle filter with importance sampling. Details about Kalman filter and particle filter can be found in Appendix~\ref{appendix:BF}. We finish this section by giving three examples of the structural mixed transition model.

\subsection{Examples of transition models}

For illustration, we list below three examples for the two transition models with probit and logistic response functions. Especially the transition model with a probit response function will be used in a numerical analysis in Section~\ref{section:numerics}. Note that, in these examples, no cure event is modelled\footnote{In practice, the cure event is usually included in the Loss-given-Default (LGD) modelling.}. Namely the default rating is absorbing and hence there is no transitions from default to the (non-default) performing ratings. Therefore, the last row of the transition matrix is trivial and does not need to be modelled. 

\subsubsection{One-factor default-only model with probit response function}\label{subsubsection:example1}

The default-only model describes the migrations by only separating default and non-default events for each rating (group), and omits the migrations between different performing (i.e., non-default) ratings. Therefore in such model only default events, and hence the probability of default (PD), of different performing ratings are modelled. This type of model is usually used to calibrate to the defaults on rating level and then extended to be able to simulate the rating migrations when necessary. The default behaviour is modelled using a one dimensional latent driving factor $x_k$ at period $k$ with probit response function, i.e., $g=\Phi$, for $\Phi$ being the cumulative distribution function of the standard normal distribution. In order to keep the PD  monotone with the ratings, i.e.\ a worse rating always has a higher PD, all different PD (signals) have the same sensitivity to the common factors. Hence the parameters $K$ and $L$ in \eqref{eq:model-m-theta} reduce to scalars. In this case the one-factor default only model, for $ k=1,\ldots, n$, and $i=1, \ldots, R-1$, is given by 
\begin{equation}\label{eq:probitT_DO}
\begin{aligned}
\theta_{i, k} &= d_i + Kx_{k} + Lu_{k}, \\
T_{iR, k} &= \pd_{i, k} = \Phi(\theta_{i, k}),\\
\log p(m_{i, k}|\theta_{i, k}) &= m_{iR,k}\log \pd_{i, k}+ (N_{i, k}- m_{iR,k})\log (1-\pd_{i, k}) + \log \binom{N_{i, k}}{m_{iR,k}}\,,\\
\end{aligned}
\end{equation}
where $\binom{N_{i, k}}{m_{iR,k}}$ is the binomial coefficient. When $R=2$, i.e.\ the whole portfolio is only separated by default and non-default states, this model belongs to the framework of the famous Merton model, see \citet{merton1974pricing}. 

\subsubsection{Two-factor model with probit response function}\label{subsubsection:example2}

This model is an extension of the one-factor default-only model in Section~\ref{subsubsection:example1}. In this model, the transitions between the non-default (performing) ratings are also modelled. All the transitions are divided into two parts, default migrations and performing migrations. The default migrations describe the rating transitions from a performing rating to default, while the performing migrations describe the transitions among the performing ratings. Each part, i.e.\ the default migration or the performing migration, is modelled by a (discrete time) stochastic process. Similar to the one-factor default-only model in Section~\ref{subsubsection:example1}, the sensitivity parameters $K$ and $L$ in Equation~\eqref{eq:model-m-theta} are set to be the same for different default migrations and different performing migrations to keep  feasibility (i.e., the transition probabilities should always be larger than zero) and the monotonicity of the PD. Consequently, the parameters $K$ and $L$ become two-by-two diagonal matrices, in which each diagonal element is the sensitivity of the default or performing signal to its corresponding common factor. The response function $g$ is chosen to be a probit function. Note that, in order to make sure that each row of the transition matrix sums up to $1$, the cumulative transition probability is actually described through the response function. Therefore, the model is described as, for $ k=1,\ldots, n$ and for $i,j =1,\ldots, R-1$,
\begin{equation}\label{eq:probitT}
\begin{aligned}
x_{k} &= [x_{k}^{(D)}, x_{k}^{(P)}]^\top, \\
u_{k} &= [u_{k}^{(D)}, u_{k}^{(P)}]^\top, \\
\theta_{k} &= [\theta^{(D)}_{k}, \theta^{(P)}_{k}]^\top = Kx_{k} + Lu_{k}\,, \quad K = \diag(k_d, k_p) \,, \quad L = \diag(l_d, l_p)\\
T_{iR, k} &= \Phi(d_{i, R} + \theta^{(D)}_{k})\,, \\
T_{ij,k} &=  (1-T_{iR, k})T_{ij, k}^{ND}\,,\\
T_{ij, k}^{ND} &= \left((\Phi(d_{i,j}+\theta^{(P)}_{k})-\Phi( d_{i,j+1}+\theta^{(P)}_{k})\right)\,, \quad 
 d_{i, R}=-\infty, \\
\log p(m_{i, k}|\theta_{i, k}) &= \sum_{j=1}^{R} m_{ij,k}\log T_{ij,k}+ \log \frac{N_{i, k}!}{\prod_{j=1}^{R} m_{ij,k}!}\,,
\end{aligned}
\end{equation}
where the $T_{ij, k}^{ND}$ are the migration probabilities given no default at period $k$.  The two-factor model can be  extended to multifactor models, see the next section. 

\subsubsection{Multi-factor model with logistic response function}
The standard logistic function is given by $p(x) = \frac{1}{1+e^{-x}} = \frac{e^{x}}{1+e^{x}}$. Extending this function to the multi-variate case and apply it as the response function $g$ in model~\eqref{eq:model-m-theta}, one can build a multi-factor model based on the logistic function as follows. For $k=1,\ldots, n\,$, $i=1,\ldots R-1$, $j =1,\ldots R$,
\begin{equation}\label{eq:logitT}
\begin{aligned}
T_{ij,k} &=  \frac{\exp(\theta_{ij,k})}{\sum_{j=1}^R \exp(\theta_{ij,k})}\,,\\
\log p(m_{i, k}|\theta_{i, k})   
&= \sum_{j=1}^R m_{ij, k}(\theta_{ij,k})  - N_{i,k} \log\big(\sum_{j=1}^R \exp(\theta_{ij,k})\big) \\
& \qquad  + \log\frac{N_{i, k}!}{\prod_{j=1}^{R} m_{ij,k}!}\,.
\end{aligned}
\end{equation}

\section{Calibration of the transition model}\label{section:calibration}

In this section we presents the ML approach to estimate the parameters (to be specified) in the transition model. Note that analytical formulae for the likelihood function of the observed migrations w.r.t.\ the model parameters, see Equation~\eqref{eq:likelihood}, are not available due to the non-Gaussian and non-linear nature of the transition model. Therefore, approximations are needed. We first propose a Laplace approximation of the likelihood. This approximation is very accurate when modeling high default portfolios with a relatively large number of observations, such as retail portfolios. The advantage of the Laplace approximation is that it produces a likelihood function that can be analytically computed by using the Kalman filter. Therefore, ML estimation can be easily conducted by using numerical optimization. 
However, for low default portfolios with a smaller number of clients, such as non-retail portfolios, the accuracy of the Laplace approximation is not assured. Consequently, the resulting ML estimator can be severely biased. To circumvent this problem, for a low default portfolio, we design a particle filter algorithm to estimate the likelihood function. This algorithm involves an importance sampling approach which makes the particle filter estimation very efficient. One of the biggest problems of using a particle filter to approximate the likelihood is that the likelihood function is not continuous w.r.t.\ the model parameters. Consequently, usual numerical optimization is not feasible in this case to obtain the ML estimates. As a remedy, we propose a grid-based approach using Gaussian process regression (GPR).

\subsection{Approach for high default portfolios: Laplace approximation for the likelihood function}\label{subsection:MLE_Lapalce}

This subsection describes the Laplace approximation of the likelihood function and the corresponding algorithm to obtain the ML estimates of the unknown model parameters $\psi$ representing $d$, $K$, $L$, $A$ and $Q$ in the transition model \eqref{eq:model-m-theta} given the observed migrations $M$. Recall Equation~\eqref{eq:calibration-psi}, we apply Laplace's method to the integrals, i.e.\ we use a second order Taylor expansion of $p(M|\theta)$ around the mode $\tilde{\theta}$ of $p(M|\theta)$. The estimation of the mode uses the approach in \citet{durbin2012time}. We will shortly explain this in Section~\ref{subsection:mode}. Here, for the time being, we suppose the mode $\tilde{\theta}$ is known to us. Denote 
\begin{align}\label{eq:hatD-hatH}
\hat{D}(\theta_0) = \frac{\partial \log p(M|\theta)}{\partial \theta}|_{\theta = \theta_0}\,, \qquad 
\hat{H}(\theta_0) = \frac{\partial^2\log p(M|\theta)}{\partial \theta \partial \theta^T}|_{\theta = \theta_0}\,.
\end{align}

Note that, use \eqref{eq:M_theta},
\begin{equation*}
\begin{aligned}
\log p(M|\theta) &= \sum_{k=1}^{n} \log p(M_k|\theta_k) \\
&= \sum_{k=1}^{n}\sum_{i=1}^{R-1} \log p(m_{i, k}|\theta_{i, k}).
\end{aligned}
\end{equation*}
Therefore, we have
\begin{equation}\label{eq:laplace_DH}
\begin{aligned}
\hat{D}(\theta) & =: \frac{\partial \log p(M|\theta)}{\partial \theta} = \big[\frac{\partial \log p(m_{1,1}|\theta_{1, 1})}{\partial \theta_{1, 1}}, \ldots, \frac{\partial \log p(m_{R-1,n}|\theta_{R-1,n})}{\partial \theta_{R-1,n}}\big], \\
\hat{H}(\theta) & =: \frac{\partial^2\log p(M|\theta)}{\partial \theta \partial \theta^T} = \mbox{blkdiag}\big(\frac{\partial^2\log p(m_{1,1}|\theta_{1, 1})}{\partial \theta_{1, 1} \partial \theta_{1, 1}^T}, \ldots, \frac{\partial^2\log p(m_{R-1,n}|\theta_{R-1,n})}{\partial \theta_{R-1,n} \partial \theta_{R-1,n}^T}\big)\,,
\end{aligned}
\end{equation}
where blkdiag refers to a block diagonal matrix. Then using a second order Taylor expansion around the model $\tilde{\theta}$, one obtains, recalling \eqref{eq:calibration-psi},
\begin{align*}
p(M|\psi) 
& =        \int p(M|\theta)p(\theta|\psi)\,\dd\theta \\
            & =        \int e^{\log p(M|\theta)}p(\theta|\psi)\, \dd\theta \\
            &\approx  \int \exp\big(\log p(M|\tilde{\theta})+\hat{D}(\tilde{\theta})^\top(\theta-\tilde{\theta})+\frac{1}{2}(\theta-\tilde{\theta})^\top\hat{H}(\tilde{\theta})(\theta-\tilde{\theta})\big)p(\theta|\psi)\, \dd\theta \\
            & =        p(M|\tilde{\theta}) \int \exp\left\{\frac{1}{2}\left(\theta-\tilde{\theta}+\hat{H}(\tilde{\theta})^{-1}\hat{D}(\tilde{\theta})\right)^\top\hat{H}(\tilde{\theta})\left(\theta-\tilde{\theta}+\hat{H}(\tilde{\theta})^{-1}\hat{D}(\tilde{\theta})\right)\right.\\
            &\left.\qquad -\frac{1}{2}\hat{D}(\tilde{\theta})^\top(\hat{H}(\tilde{\theta})^{-1})^\top\hat{D}(\tilde{\theta})\right\}p(\theta|\psi)\, \dd\theta\,. 
            \end{align*}
Let 
\[
C = (2\pi)^{n(R-1)R/2} (\det(-\hat{H}(\tilde{\theta})^{-1}))^{1/2}p(M|\tilde{\theta})\exp(-\frac{1}{2}\hat{D}(\tilde{\theta})^\top(\hat{H}(\tilde{\theta})^{-1})^\top\hat{D}(\tilde{\theta}))
\]
and 
\begin{align}\label{eq:hat-M-n}
\hat{M}(\tilde{\theta})=\tilde{\theta}-\hat{H}(\tilde{\theta})^{-1}\hat{D}(\tilde{\theta})\,.
\end{align}
With 
\begin{align}\label{eq:mntheta2}
p(\hat{M}(\tilde{\theta})|\theta) &= (2\pi)^{-n(R-1)R/2} (\det(-\hat{H}(\tilde{\theta})^{-1}))^{-1/2}                             \exp({\frac{1}{2}(\theta-\hat{M}(\tilde{\theta}))^\top\hat{H}(\tilde{\theta})(\theta-\hat{M}(\tilde{\theta}))})
\end{align}
and in view of \eqref{eq:calibration-psi} for $\hat{M}(\tilde{\theta})$ instead of $\theta$,
it then holds
\begin{align}\label{eq:laplace}       
p(M|\psi) 
  &\approx          
      C\int p(\hat{M}(\tilde{\theta})|\theta)p(\theta|\psi)\,\dd\theta\nonumber\\
  & =        C\,p(\hat{M}(\tilde{\theta})|\psi)\,. 
\end{align}
Note that \eqref{eq:mntheta2} expresses that, conditional on $\theta$, $\hat{M}(\tilde{\theta})$ has a $N(\theta, -\hat{H}(\tilde{\theta})^{-1})$ distribution, using that $\hat{H}(\tilde{\theta})$ is negative definite since $\tilde\theta$ is the mode. According to Equation~\eqref{eq:laplace_DH}, 
denote $\hat{D}(\tilde{\theta}) = [\hat{D}_{1, 1}(\tilde{\theta}_{1, 1}), \ldots, \hat{D}_{R-1, n}(\tilde{\theta})]$ and $\hat{H}(\tilde{\theta}) = \mbox{blkdiag}(\hat{H}_{1, 1}(\tilde{\theta}_{1, 1}), \ldots, \hat{H}_{R-1, n}(\tilde{\theta}_{R-1, n}))$. One can formulate a linear Gaussian state space model for $\hat{M}(\tilde{\theta})=\{\hat{M}_1(\tilde{\theta}_{1}), \ldots, \hat{M}_n(\tilde{\theta}_{n})\}$, with the matrix $\hat{M}_k(\tilde{\theta}_{k})=(\hat{m}_{ij, k}, i=1, \ldots, R-1, j=1, \ldots, R) , k=1, \ldots, n$ :
\begin{equation}\label{eq:SSM_M}
\begin{aligned}
\hat{m}_{ij,k} &= \theta_{ij, k} + \epsilon_k = d_{i,j} + K_{i, j}^\top x_k +  L_{i, j}^\top u_{k} + \epsilon_{ij, k},\\
x_k &= Ax_{k-1} + \eta_k, \qquad \eta_k\sim N(0, Q), 
\end{aligned}
\end{equation}
with $\epsilon_{i, k} = [\epsilon_{i1, k}, \ldots, \epsilon_{iR, k}]\sim N(0, -\hat{H}(\tilde{\theta})^{-1}$) and the likelihood $p(\hat{M}(\tilde{\theta})|\psi)$ can be computed by Kalman filtering as described in Section~\ref{subsection:KF}. 

\subsubsection{Mode estimation}\label{subsection:mode}

The Taylor expansion applied to the density $p(M|\theta)$ is around the mode $\tilde{\theta}$ of the posterior density $p(\theta|M)$, i.e.
\begin{equation*}
\tilde{\theta} = \mathop{\argmax}_{\theta} p(\theta|M).
\end{equation*} 
The posterior density $p(\theta|M)$ does not have an explicit expression from which we
can obtain the mode analytically. Therefore, a Newton-Raphson method is applied to numerically maximize $p(\theta|M)$ w.r.t.\  $\theta$, see \citet{durbin2012time}. Suppose the Newton-Raphson estimate at the current step is $\hat{\theta}$, the Newton-Raphson update of the current estimate $\hat{\theta}$, denoted by $\hat{\theta}^+$,  is expressed by
\begin{equation}\label{eq:NR_update}
\hat{\theta}^+ = \hat{\theta} - [\ddot{p}(\theta|M)|_{\theta=\hat{\theta}}]^{-1}\dot{p}(\theta|M)|_{\theta=\hat{\theta}},
\end{equation}
where the dots denote differentiation w.r.t.\ $\theta$.
Note that (in self-evident notation for densities)
\begin{equation}\label{eq:logp_signal}
\log p(\theta|M) = \log p(M|\theta) + \log p(\theta) - \log p(M). 
\end{equation}
The explicit expression for the gradients $\dot{p}(\theta|M)$ and Hessian $\ddot{p}(\theta|M)$ can be  derived according to Equation~\eqref{eq:logp_signal}, since the analytic formulas for the conditional likelihood $p(M|\theta)$ and the marginal density $p(\theta)$ are available. Specially the gradient and Hessian of $p(M|\theta)$ are computed as $\hat{D}(\hat{\theta})$ and $\hat{H}(\hat{\theta})$ in Equation~\eqref{eq:laplace_DH} respectively.  Moreover, \citet{durbin2012time} shows that the Newton-Raphson update $\hat{\theta}^+$ in Equation~\eqref{eq:NR_update} is equal to the posterior estimates of $\theta_1, \ldots, \theta_n$ in the linear Gaussian state space model~\eqref{eq:SSM_M}. These posterior estimates can  be computed through the Kalman filter algorithm in Section~\ref{subsection:KF}. 

\subsubsection{Algorithms for mode estimation and MLE based on Laplace approximation}

In this section we present the algorithm for the mode estimation and MLE based on the Laplace approximation.

\begin{algorithm}[{\bf Mode estimation (Newton-Raphson)}]\label{algo:KF_mode}
\mbox{}
Given are the parameters $d$, $K$, $L$, and the process $u_i$, $i=1,\ldots,n$, see \eqref{eq:migration}.
\begin{description}
\item[Initialization:] Define the initial guess $\hat{\theta}^+$ for the mode of $p(\theta|M)$. Define the threshold $C_T$ as the condition for judging the convergence of Newton-Raphson algorithm. Define the initial $\hat{\theta}$ such that $\lVert \hat{\theta}^+-\hat{\theta} \rVert > C_T$
\item[Iteration:] While $\lVert \hat{\theta}^+-\hat{\theta} \rVert > C_T$, 
\begin{enumerate}
\item Reset $\hat{\theta} = \hat{\theta}^+$.
\item Compute the gradient $\hat{D}(\hat{\theta})$ and Hessian $\hat{H}(\hat{\theta})$ in Equation~\eqref{eq:laplace_DH}.
\item Run the Kalman filter Algorithm~\ref{algo:KF} on model~\eqref{eq:SSM_M}.
The output of the Kalman filter is $\hat{x}_k$, $k=1,\ldots,n$.
\item Update $\hat{\theta}^+$ to $(d + K\hat{x}_1 + Lu_1, \ldots, d + K\hat{x}_n + Lu_n)$. 
\end{enumerate}
\item[Convergence:] Once $\lVert \hat{\theta}^+-\hat{\theta} \rVert \leq C_T$, the mode estimation is given by $\hat{\theta}^+$.
\end{description}
\end{algorithm}

\begin{algorithm}[{\bf MLE (Laplace approximation)}]\label{algo:MLE_laplace}
\mbox{}\vspace{-0.5em}
\begin{description}
\item[Likelihood function:] Use the parameter value $\psi$ as input, build the likelihood function $p(M|\psi)$ by:
\begin{enumerate}
\item Run the mode estimation Algorithm~\ref{algo:KF_mode} and obtain the mode estimate $\tilde{\theta}_\psi$.
\item Given the mode estimate $\tilde{\theta}_\psi$, compute the gradient $\hat{D}(\tilde{\theta}_\psi)$ and Hessian $\hat{H}(\tilde{\theta}_\psi)$.
\item Compute the likelihood $p(\hat{M}(\tilde{\theta}_\psi)|\psi)$ in Equation~\eqref{eq:laplace} by applying the Kalman filter to model~\eqref{eq:SSM_M}.
\item Approximate the likelihood function by the Laplace approximation of \eqref{eq:laplace}, i.e.\ $p(M|\psi)\approx Cp(\hat{M}(\tilde{\theta}_\psi)|\psi)$.
\end{enumerate}

\item[Maximization:] Maximize the likelihood function $p(\hat{M}(\tilde{\theta}_\psi)|\psi)$ and obtain the optimal parameter $\tilde{\psi}$, i.e.
\begin{align*}
\tilde{\psi} = \mathop{\argmax}_{\psi} Cp(\hat{M}(\tilde{\theta}_\psi)|\psi)\,.
\end{align*}

\item [Latent states estimation:] Given $\tilde{\psi}$, run the mode estimation Algorithm~\ref{algo:KF_mode} and obtain the estimates $\hat{x}_k$ of the latent process $x_k$, $k=1,\ldots, n$.
\end{description}
\end{algorithm}

\subsection{Approach for low default portfolios: particle filter for the likelihood function}\label{subsection:smc}

The proposed Laplace approach applied to the likelihood requires a sufficiently large number of observed migrations to ensure a satisfactory degree of accuracy. However, this property is not always met. For instance, when modeling the default events of large corporate and financial institutions, the observed numbers of default are very limited since these companies rarely default. In this case, we propose a particle filter, see e.g.~\citet{arulampalam2002tutorial}, \citet{cappe2007overview}, \citet{doucet2009tutorial} or \citet{he2021kalman},  based algorithm to estimate the likelihood function. We design an importance sampling approach that leverages on the Laplace approximation, so that the particle filter becomes much more efficient and can  deal with the outliers in the data. 

\subsubsection{Importance sampling}\label{subsection:PF_is}

In Section~\ref{subsection:PF} of the Appendix the basic (bootstrap) particle filter is presented. This filter will be used to estimate the likelihood of the observations $p(M|\psi)$. However, the particle filter is very inefficient when outliers in the data are present. Experiments, see for instance Section~\ref{subsection:assess1}, show that the likelihood function obtained from the particle filter is very sensitive to outliers. Therefore, a large number of Monte Carlo samples need  to be generated to obtain an accurate estimate of the posterior distribution of the outlier and consequently to obtain also an accurate estimate of the likelihood. In this case, importance sampling is used to enhance the performance of the particle filter. The idea of importance sampling is to sample the Monte Carlo particles from a proposed density that is as close as possible to the posterior density $p(x_k|M_{1:k})$, where $(x_k)_{k=1, \ldots, n}$ is the latent process and $(M_k)_{k1, \ldots, n}$ is the migration matrix sequence. This proposed density is the so-called importance density. Denote the importance density by $q(x_k|M_{1:k}, x_{k-1})$. In the present context Equation~\eqref{eq:posterior} takes the form
\begin{equation}\label{eq:posterior_is}
\begin{aligned}
p(x_k|M_{1:k}, \psi) &= \frac{p(M_k\mid x_k,M_{1:k-1}, \psi)\int p(x_k\mid x_{k-1}, \psi)p(x_{k-1}\mid M_{1:k-1}, \psi)\, \ud x_{k-1}}{p(M_k\mid M_{1:k-1}, \psi)}\\
& = \frac{\int \alpha(x_k, x_{k-1}, M_{1:k}\mid \psi) q(x_k|M_{1:k}, x_{k-1}) p(x_{k-1}\mid M_{1:k-1}, \psi)\, \ud x_{k-1}}{p(M_k\mid M_{1:k-1}, \psi)}\,,
\end{aligned}
\end{equation}                                      
where
\begin{equation}\label{eq:importance_weight}
   \alpha(x_k, x_{k-1}, M_{1:k}\mid \psi)=\frac{p(M_k\mid x_k,M_{1:k-1}, \psi)p(x_k\mid x_{k-1}, \psi)}{q(x_k|M_{1:k}, x_{k-1})} 
\end{equation}
is the so-called importance weight. Consequently, the conditional likelihood in Equation~\ref{eq:cond_l} is reformulated as
\begin{equation}\label{eq:cond_l_is}
p(M_k\mid M_{1:k-1}, \psi)= \int\! \int \alpha(x_k, x_{k-1}, M_{1:k}\mid \psi) q(x_k|M_{1:k}, x_{k-1}) p(x_{k-1}\mid M_{1:k-1}, \psi)\, \ud x_{k-1} \ud x_{k}.
\end{equation}
When implementing importance sampling in the particle filter, according to Equation~\eqref{eq:posterior_is} and~\eqref{eq:cond_l_is}, the particles of $x_k$ are generated from the importance density $q(x_k|M_{1:k}, x_{k-1})$, instead of the transition density $p(x_k\mid x_{k-1})$ as described in Section~\ref{subsection:PF}. 
As a result, according to Equation~\eqref{eq:PF_nosampling}, the formula for the weights in Equation~\eqref{eq:w_PF} now becomes
\begin{equation}\label{eq:w_PF2}
w_{k}^{(i)} = \frac{w_{k-1}^{(i)}\alpha(x_k^{(i)}, x_{k-1}^{(i)}, M_{1:k}\mid \psi)}{\sum_{i=1}^N w_{k-1}^{(i)}\alpha(x_k^{(i)}, x_{k-1}^{(i)}, M_{1:k}\mid \psi)}\,.
\end{equation}
The biggest difficulty when implementing importance sampling is to find a proper importance density. To this end, we propose an importance density by leveraging on the Laplace approximation described in Section~\ref{subsection:MLE_Lapalce}. Once the approximation $\tilde{\theta}$ of the mode is obtained, the Laplace approximation is applied to the density $p(M|\theta)$. The approximated density $p(\hat{M}(\tilde{\theta})|\theta)$ results in a linear Gaussian state space model as in Equation~\eqref{eq:SSM_M}. Therefore, by running the Kalman filter algorithm on the model~\eqref{eq:SSM_M} one obtains an estimate of the posterior distribution of $x_k$. 

To summarize, we present here an alternative algorithm to Algorithm~\ref{algo:MLE_laplace}. This alternative algorithm is specially designed for the low default portfolio, in which the number of observations is not large enough to support a good approximation using Laplace approach in Algorithm~\ref{algo:MLE_laplace}.  Note that we will need the mode estimation procedure as in Algorithm~\ref{algo:KF_mode}. Recall  the notation $M_{1:k} =M_1, \ldots, M_k$. When $k=0$, then this is an empty sequence.
\begin{algorithm}[{\bf Particle filter with importance sampling for given parameter $\psi$}]\label{algo:PF_is}
Given is the parameter vector $\psi$.
\begin{description}
\item[Initialization:] 
Assume an initial distribution $p(x_{0})$ for the latent states $x_0$, and sample from this initial distribution to get the $N$ particles $\{x_{0}^{(i)}, i=1,\ldots,N\}$ and define $\{w_{0}^{(i)}=\frac{1}{N}, i=1,\ldots, N\}$.
\item [Importance density:] Run Algorithm~\ref{algo:KF_mode} and obtain the mode estimation of $\theta$. Given the mode $\tilde{\theta}$, run the Kalman filter Algorithm~\eqref{algo:KF} on model~\eqref{eq:SSM_M} to obtain the posterior distribution of the latent process $\{x_k, k=1, \ldots, n\}$. These posterior distributions, denoted by $q(x_k|M_{1:k}, x_{k-1}^{(i)}, \psi), k=1, \ldots, n$, are used as the importance densities.
\item [Particle filter recursion:] For $k=1, \ldots, n$, suppose the estimated posterior distribution at time $k-1$ is given by $p(x_{k-1}|M_{1:k-1}, \psi) \approx \frac{1}{N}\sum_{i=1}^N \delta_{x_{k-1}^{(i)}}(x_{k-1})$.
\begin{enumerate}
\item Sample $N$ particles from the importance density, i.e.\ $\tilde{x}_{k}^{(i)} \sim q(x_k|M_{1:k}, x_{k-1}^{(i)}, \psi)$, for $i=1,\ldots,N$.
\item Compute the weights according to Equations~\eqref{eq:importance_weight} and~\eqref{eq:w_PF2}, namely 
\begin{align*}
w_{k}^{(i)} = \frac{w_{k-1}^{(i)}\alpha(\tilde{x}_k^{(i)}, x_{k-1}^{(i)}, M_{1:k}\mid \psi)}{\sum_{i=1}^N w_{k-1}^{(i)}\alpha(\tilde{x}_k^{(i)}, x_{k-1}^{(i)}, M_{1:k}\mid \psi)},
\end{align*}
with the initial  $w_0$ as specified in the Initialization step, and the other $w_k$ are calculated from the iteration, using the equation above with
\begin{align*}
\alpha(\tilde{x}_k^{(i)}, x_{k-1}^{(i)}, M_{1:k}\mid \psi)=\frac{p(M_k\mid \tilde{x}_k^{(i)},M_{1:k-1}, \psi)p(\tilde{x}_k^{(i)}\mid x_{k-1}^{(i)}, \psi)}{q(\tilde{x}_k^{(i)}|M_{1:k}, x_{k-1}^{(i)}, \psi)}),
\end{align*}
\item Resample $\{x_{k}^{(1)}, \ldots, x_{k}^{(N)}\}$ from $\{\tilde{x}_{k}^{(1)}, \ldots, \tilde{x}_{k}^{(N)}\}$ with probabilities $\{w_k^{(1)}, \ldots, w_k^{(N)}\}$.
\item The posterior distribution and the conditional likelihood at time $k$ are estimated as 
\begin{equation}\label{eq:PF_cond_l}
\begin{aligned}
p(x_{k}|M_{1:k}, \psi) &\approx \frac{1}{N}\sum_{i=1}^N \delta_{\tilde{x}_k^{(i)}}(x_k), \\
p(M_{k}|M_{1:k-1}, \psi) &\approx \frac{1}{N} \sum_{i=1}^N  p(M_k|x_k^{(i)}, \psi).
\end{aligned}
\end{equation}
\end{enumerate}
\item [Likelihood:] Compute the log-likelihood of the observation as
\begin{align*}
\log p(M|\psi)\approx \sum_{k=1}^n \log p(M_{k}|M_{1:k-1}, \psi)\,.
\end{align*}
\end{description}
\end{algorithm}

\subsubsection{MLE for the particle filter}

The results in Section~\ref{subsection:PF_is} show that for a given parameter $\psi$, the (conditional) likelihood $p(M_k|M_{1:k-1}, \psi)$ can be estimated according to Equation~\eqref{eq:PF_cond_l}. However, the particle filter estimate of the conditional likelihood is not a continuous function of  $\psi$. This is due to the resampling steps. At each time $k$, in multinomial resampling, a piecewise constant and hence discontinuous cumulative distribution function is defined by the weights $\{w_k^{(i)}, i=1, \ldots, N\}$ and particles $\{x_k^{(i)}, i=1, \ldots, N\}$. A small change in the parameter $\psi$ will cause a small change in the importance weights $\{w_k^{(i)}\}$. But, due to the discontinuous character of the multinomial cumulative distribution function, a small change in the importance weights will potentially generate a different set of resampled particles $\{x_k^{(i)}\}$ and hence the likelihood function estimate will not be continuous in $\psi$. This discontinuity problem in the static parameter estimation using a particle filter has generated a lot of interest over the past few years and many techniques have been proposed to solve it, for instance gradient approach of \citet{gradient2005}, expectation maximization (EM) in \citet{Partile2004} and \citet{ParaEM2008}, smoothing likelihood in \citet{Pitt2002}, and Markov Chain Monte Carlo (MCMC) in \citet{Chopin2011}. However, these approaches have difficulties in terms of implementation, see for instance \citet{KANTAS2009774}.  For instance, in the gradient approach the step size is difficult to tune; the EM method requires a large number of particles to ensure the likelihood in the expectation step to increase monotonically; and the MCMC approach is usually time consuming. Therefore, in this paper we propose a smoothing likelihood method based on Gaussian process regression (GPR) to calibrate the model parameters, and the experiments in Section \ref{subsection:assess3} show that the GPR based MLE approach is robust and fast.

\subsubsection{Smooth likelihood function by GPR}

The idea of the smoothing likelihood approach is to approximate the likelihood function by a smooth function, see \citet{Pitt2002}. In this paper, we choose to use GPR to approximate the likelihood function. A short introduction to GPR is presented in the Appendix, Section~\ref{appendix:GPR}. Before applying the GPR methodology to the ML estimation, it needs to be trained to evaluate the likelihood. Given a set of pre-selected model parameters, the likelihoods are approximated by using the proposed particle filter Algorithm~\ref{algo:PF_is}. The pre-selected model parameters and the corresponding likelihoods are used to construct the grid for the model parameters. Then the GPR is applied to fit the grid. Once the GPR is trained, the ML algorithm can be applied and one obtains an approximation of the ML estimator. The algorithm is summarized as follows.
\begin{algorithm}[{\bf PF with GPR}]\label{algo:SMC_GPR}
\mbox{}
\begin{enumerate}
\item Define the grid for the value of model parameters.
\item Given each point (a vector of model parameter values) on the grid, run Algorithm~\ref{algo:PF_is} with the proposed importance sampling to obtain the corresponding  log-likelihood.
\item Train the GPR on the grid with the corresponding log-likelihood. The trained GPR can be defined as a function between the model parameters and the log-likelihood, denoted by $f_{GPR}(\psi)$.
\item Find the estimator $\hat{\psi}$ such that 
\begin{equation*}
\hat{\psi} = \mathop{\argmax}_{\psi} f_{GPR}(\psi).
\end{equation*} 
This $\hat{\psi}$ is the approximate ML estimator. 
\end{enumerate}

\end{algorithm}

\section{Numerical results}\label{section:numerics}

\subsection{Setup for the transition model}\label{subsection:TM_model_setup}

We use the one factor default-only model and the two factor transition model with the probit response function, see Equations~\eqref{eq:model-m-theta}, \eqref{eq:probitT_DO} and~\eqref{eq:probitT}, as the examples to illustrate the performance of the proposed calibration methodologies. Specially, the default-only model is used to show the likelihood function approximation of the proposed Laplace and the particle filter (with importance sampling) approximation, see Section~\ref{subsection:assess1}. It is also used, in Section~\ref{subsection:assess3}, to illustrate the calibration performance of the particle filter (PF) with the GPR approach, Algorithm~\ref{algo:SMC_GPR}. The two factor transition model is used to test the calibrations based on Laplace approximation approach, see Section~\ref{subsection:assess2}. In the experiments of this paper, we omit the observed process $(u_k)_{k=1, \ldots, n}$, i.e.\ $u_k=0$, for all $k=1, \ldots, n$ in Equation~\eqref{eq:model-m-theta}, as it does not increase the complexity from methodological perspective but just adds a few unknown model parameters of the calibration.

We use simulated migration data for the experiments, so that the `true' values of the model parameters are known and can be used as a benchmark. The simulation is done according to Equations~\eqref{eq:model-m-theta} and~\eqref{eq:probitT_DO} or~\eqref{eq:logitT}, by assuming a one- or two-factor model for the latent process $(x_k)_{k\in \mathbb{N}^+}$. The process $(x_k)_{k\in \mathbb{N}^+}$ is simulated using the Monte Carlo method based on the autoregressive model in Equation~\eqref{eq:model-m-theta}. Consequently, the migrations are simulated according to the model with Equation~\eqref{eq:probitT_DO} or~\eqref{eq:logitT}. The simulated data are supposed to have 150 time points, i.e., one assumes to observe 150 periods of migrations. We suppose to have four different ratings with three performing ratings (P1, P2 and P3) and one default rating (D). We further assume that the default state is absorbing, which means that the probabilities of the transition from D to P1, P2 or P3 are zero. The PF (with importance sampling) Algorithm~\ref{algo:PF_is}  is designed for low-default portfolios.  We assume, for the PF assessments, the long term average of the probability of defaults is  $\overline{\pd} = [0.001, 0.004, 0.01]$. The number of clients in these three ratings are assumed to be $[5000, 1000, 500]$. By contrast, the Laplace approximation is for portfolios with higher default probabilities and more number of clients. Therefore, for Laplace approximation assessments, we assume the long term average $\overline{\pd} = [0.01, 0.04, 0.1]$. The value of the $d$ parameter defined after Equation~\eqref{eq:model-m-theta} is derived from the long term average $\overline{\pd}$. The number of clients in these three ratings are assumed to be $[100000, 10000, 5000]$.
\medskip\\
Strictly speaking, the models~\eqref{eq:probitT_DO} and~\eqref{eq:probitT} have three parameters: $d$, $A$ and $K$. Specially the parameter $d$ contains quite some elements. To reduce the number of parameters, the elements of $d$, in the calibration,  we will use the auxiliary fact that
\begin{equation}\label{eq:EPhi_x}
 \mathbb{E}\left[\Phi(X)\right]= \Phi(\frac{\mu}{\sqrt{1+\sigma^2}}),   
\end{equation}
if $X\sim N(\mu, \sigma)$. To see that this relation holds, we argue as follows. We write the expectation as, with $\phi$ the standard normal density,
\[
\mathbb{E}\left[\Phi(X)\right]=\int_{-\infty}^\infty\Phi(\mu+\sigma z)\phi(z)\, \dd z = 
\int_{-\infty}^\infty\int_{-\infty}^{\mu+\sigma z}\phi(x)\phi(z)\,\dd x \dd z , 
\]
where we recognize that the double integral equals $\mathbb{P}(X\leq \mu+\sigma Z)$ with $X$ and $Z$ independent standard normals. This probability trivially equals $\mathbb{P}(X-\sigma Z\leq \mu)$. 
As $X-\sigma Z$ has a $N(0,1+\sigma^2)$ distribution, $\mathbb{P}(X-\sigma Z\leq \mu)=\Phi(\frac{\mu}{\sqrt{1+\sigma^2}})$, and we obtain \eqref{eq:EPhi_x}.

Coming back to the calibration, to reduce the number of parameters to be estimated, the elements of $d$ are estimated directly by using the average, denoted $\bar{r}$, of the observed PDs (for the default-only model~\eqref{eq:probitT_DO}) or cumulative migration probabilities (for the migration model~\eqref{eq:probitT}) and the $K$ parameter. So, we approximate the theoretical PD, according to the law of large numbers, by the sample average, i.e.\ $\bar{r}$, and
invert the relation \eqref{eq:EPhi_x} with $\mu=d$ and $\sigma=K$ to get the approximation (estimate)
\begin{equation*}
d \approx \sqrt{1+K^2}\,\Phi^{-1}(\bar{r}).
\end{equation*}
Consequently, the model parameters left to be estimated are only $A$ and $K$. 

\subsection{Laplace approximation and importance sampling}\label{subsection:assess1}

This section assesses the accuracy of the proposed Laplace and the PF approximation of the likelihood function. The likelihood profile is generated based on the simulated migration data using the one factor default-only model~\eqref{eq:probitT_DO}. The one dimensional latent process $(x_k)_{k=1, \ldots, n}$ is as described in Equation~\eqref{eq:model-m-theta} with $A=0.7$ and $Q=\sqrt{1-A^2}\approx 0.71414$. The sensitivity parameter $K$ is set to be $0.6$ for low default portfolio and $0.3$ for high default portfolio. The performance of the PF is first assessed on a low-default portfolio as described in Section~\ref{subsection:TM_model_setup}. The assessment of the PF consist of two parts: the assessment of the importance sampling and the assessment of the likelihood function. The Laplace approximation is tested on a portfolio with higher default probabilities and a higher number of clients, and the PF is used as a benchmark.

\subsubsection{Particle filter (PF) with importance sampling}

\begin{figure}[ht]
 \centering
 \includegraphics[width=0.95\textwidth]{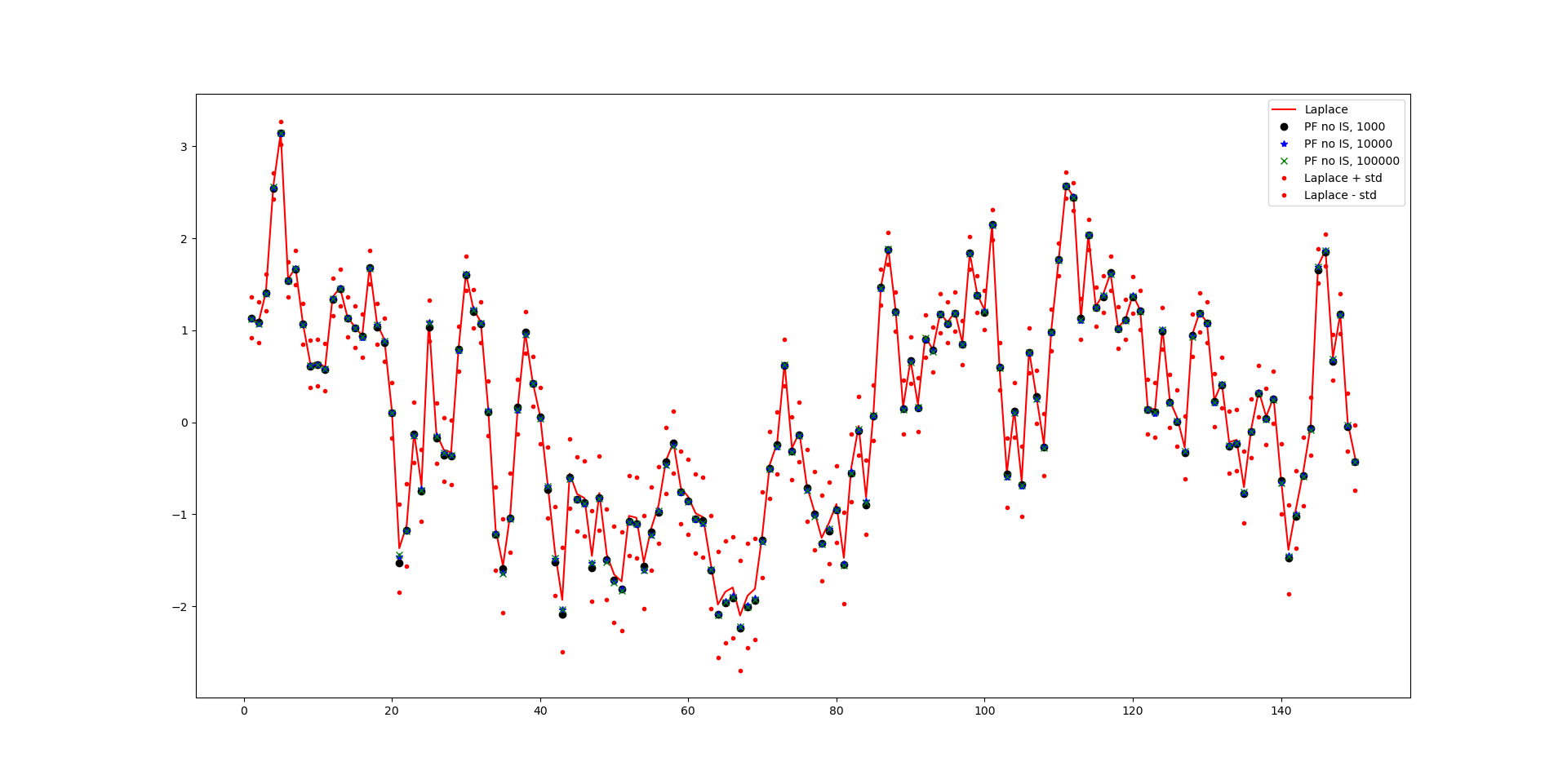}\caption{Estimated $x_k$: Laplace vs.\ PF without importance sampling. We use $A=0.7$ and $K=0.6$. The Laplace approximation and the $68\%$ confidence interval (i.e., mean $\pm$ one standard deviation) are shown in red line and dots respectively. The PF without importance sampling uses 1000, 10000, 100000 Monte Carlo samples and are shown in black dots, blue stars and green crosses, respectively. Because of the high accuracy of the procedure, the blue stars and green crosses nearly coincide with the black dots and are bearly visible.}
 \label{fig:xt_TruePara}      
\end{figure}

\begin{figure}[ht]
 \centering
 \includegraphics[width=0.95\textwidth]{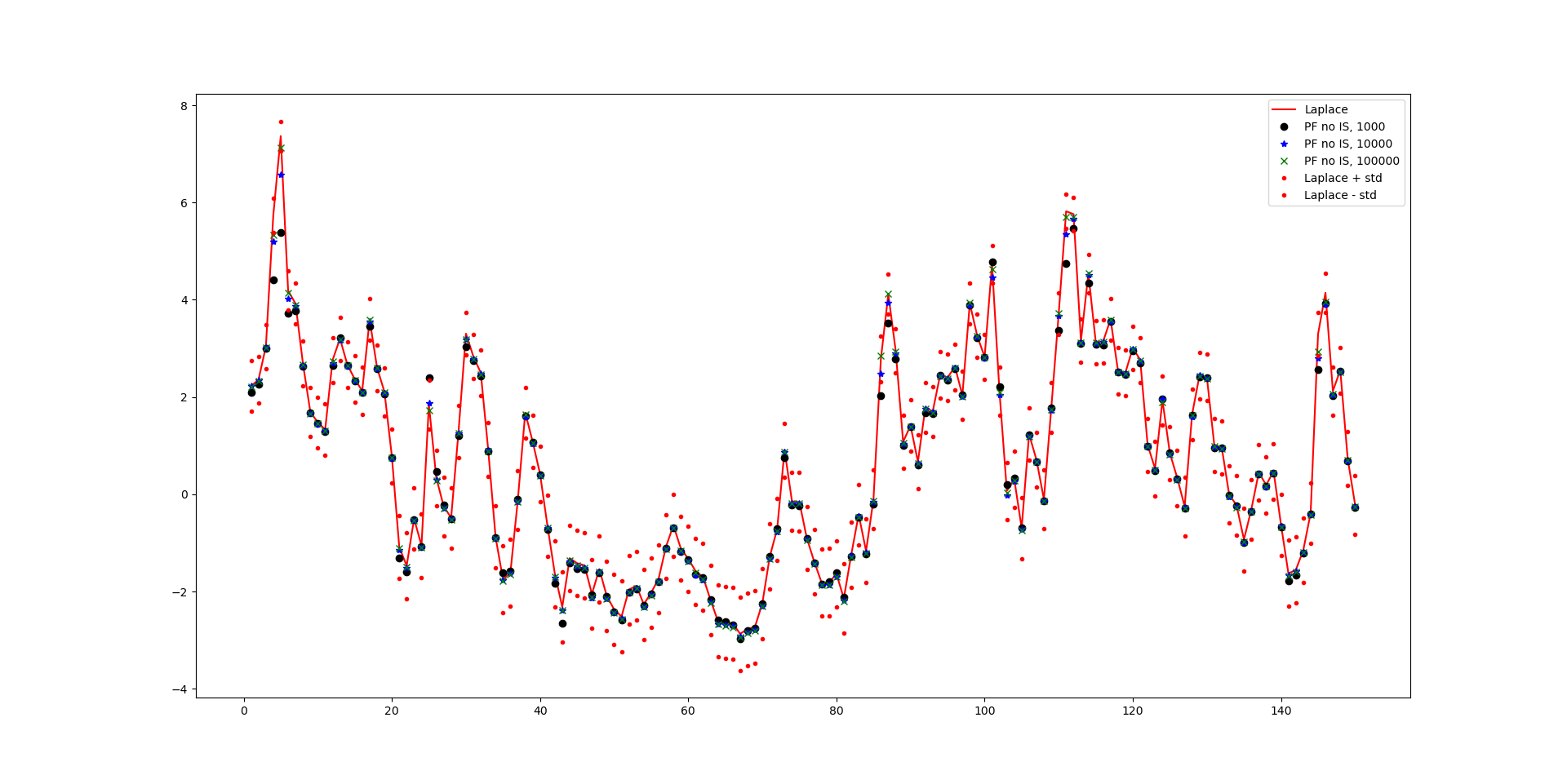}\caption{Estimated $x_k$: Laplace vs.\ PF without importance sampling. We use $A=0.7$ and $K=0.3$. The Laplace approximation and the one standard deviation confidence interval are shown in red line and dots respectively. The PF without importance sampling uses 1000, 10000, 100000 Monte Carlo samples and are shown in black dots, blue stars and green crosses, respectively.}
 \label{fig:xt_FalsePara}      
\end{figure}

The importance density used in the proposed PF algorithm is a Gaussian density with mean and covariance approximated using the Laplace approach. The approximations are obtained by running the Kalman filter algorithm on model \eqref{eq:SSM_M}. Figure~\ref{fig:xt_TruePara} and Figure~\ref{fig:xt_FalsePara} present the comparison of the estimated latent process $(x_k)_{k=1, \ldots, n}$ using Laplace approximation and PF without importance sampling. In Figure~\ref{fig:xt_TruePara} the `true' parameters are used, i.e.\ $A=0.7$ and $K=0.6$. One observes that the Laplace approximation aligns with the PF approximations. In contrast, in Figure~\ref{fig:xt_FalsePara}, the $K$ parameter is set to be $0.3$. In this case one observes that the Laplace approximation still aligns with the PF approximation with enough Monte Carlo points (100000). When the number of Monte Carlo points is not sufficiently large, one observes that the PF without importance sampling gives a biased estimation of the $x_k$. The results indicate that the Laplace approach provides a good estimate of the $x_k$ and hence can be used for importance sampling.  

\begin{figure}[ht]
 \centering
 \includegraphics[width=0.95\textwidth]{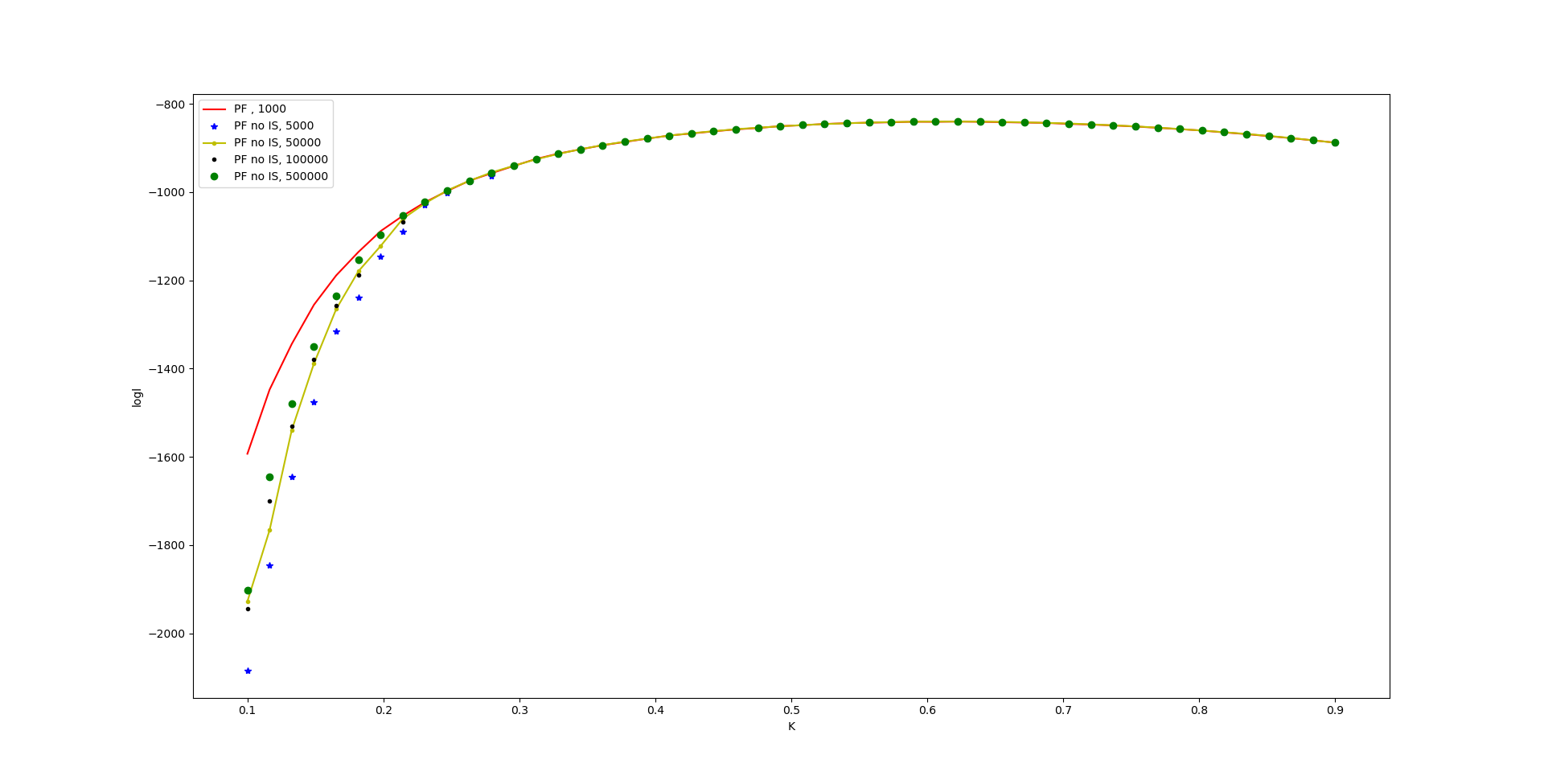}\caption{Log-likelihood profile: $K$ vs.\ log-likelihood. We use $A=0.7$. The log-likelihood profile using PF with the proposed importance sampling is shown as the red line. The log-likelihood profiles based on PF without importance sampling is used with 5000, 50000, 100000, 500000 Monte Carlo samples and are shown in blue stars, yellow dot line, black dots and green dots, respectively.}
 \label{fig:logl_k}      
\end{figure}

\begin{figure}[ht]
 \centering
 \includegraphics[width=0.95\textwidth]{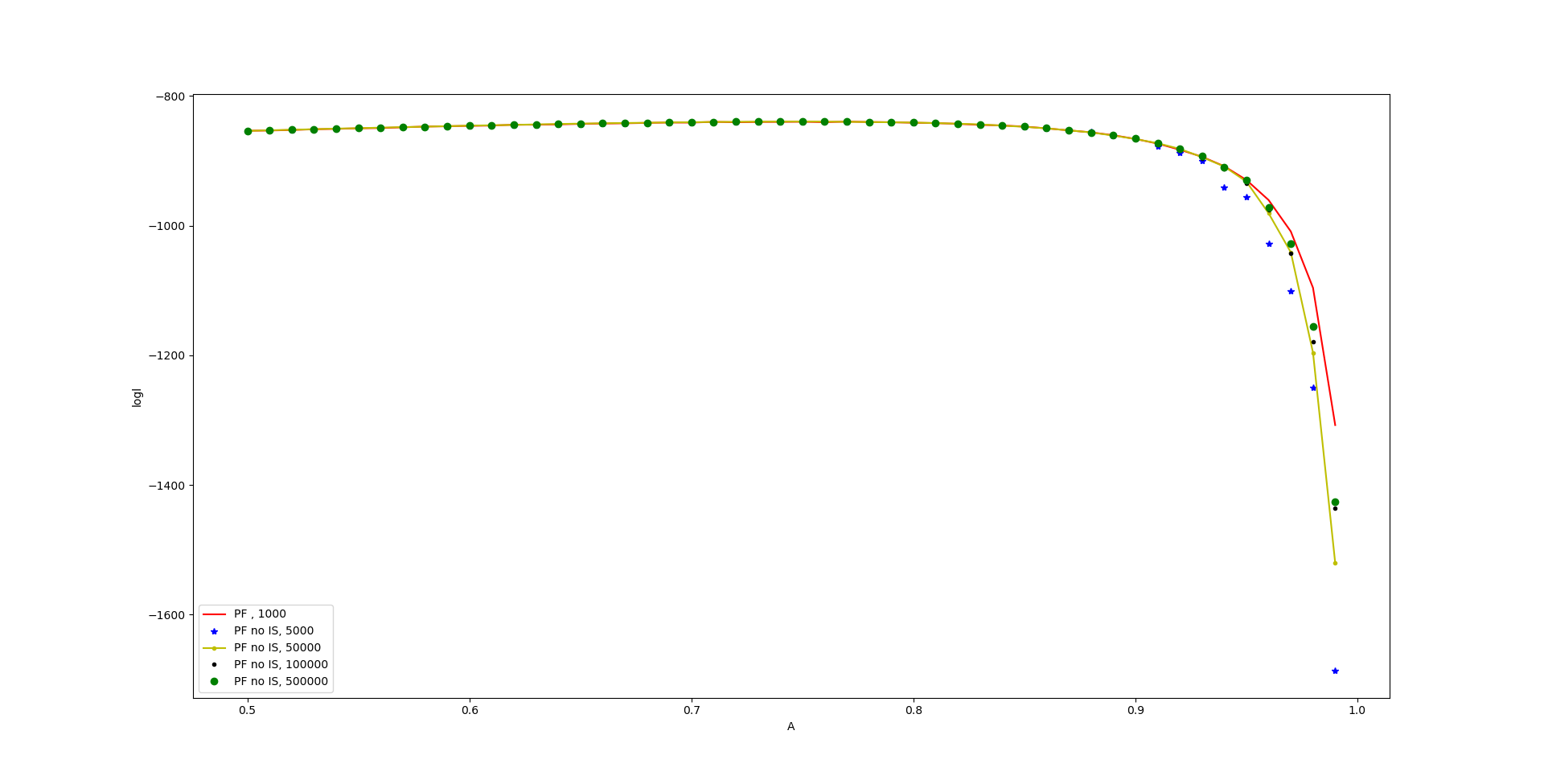}\caption{Log-likelihood profile: $A$ vs.\ log-likelihood. We use $K=0.6$. The log-likelihood profile using PF with the proposed importance sampling is shown in red. The log-likelihood profiles based on PF without importance sampling used 5000, 50000, 100000, 500000 Monte Carlo samples and are shown in blue stars, yellow dot line, black dots and green dots, respectively.}
 \label{fig:logl_a}      
\end{figure}

\noindent Figures~\ref{fig:logl_k} and \ref{fig:logl_a} show the comparison of the log-likelihood profile, with different $K$ and $A$, approximated by the PF with and without the proposed importance sampling. One observes that the PF without importance sampling converges to the PF with the importance sampling, but much slower. Especially, Figure~\ref{fig:logl_k} shows that the PF without importance sampling converges very slowly when the $K$ value is significantly smaller than the optimal (where the maximum likelihood is obtained) value. That is because the volatility of the \emph{realized} $x_k$ needs to be much larger to compensate the low value of $K$ and hence match the volatility of the signal $\theta_k$ implied from the data. Since the volatility of $x_k$ is assumed to be $Q=\sqrt{1-A^2}$,  the high volatility of $x_k$ requires the PF without importance sampling to generate extreme values for $\eta_k$ in the simulation and consequently requires a large number of Monte Carlo samples. Same reasoning applies to the cases with high value of $A$ parameters, see Figure~\ref{fig:logl_a}. When $A$ is large, the volatility of $x_k$ is small. Therefore, extreme values of $\eta_k$ need to be sampled to reach the volatility of $\theta_k$ implied from the data. This will significantly slow down the convergence speed when an appropriate importance sampling is not used.  These results suggest that the PF with the proposed importance sampling significantly increases the efficiency of the normal PF algorithm.

\subsubsection{Laplace approximation}

\begin{figure}[ht]
 \centering
 \includegraphics[width=0.80\textwidth]{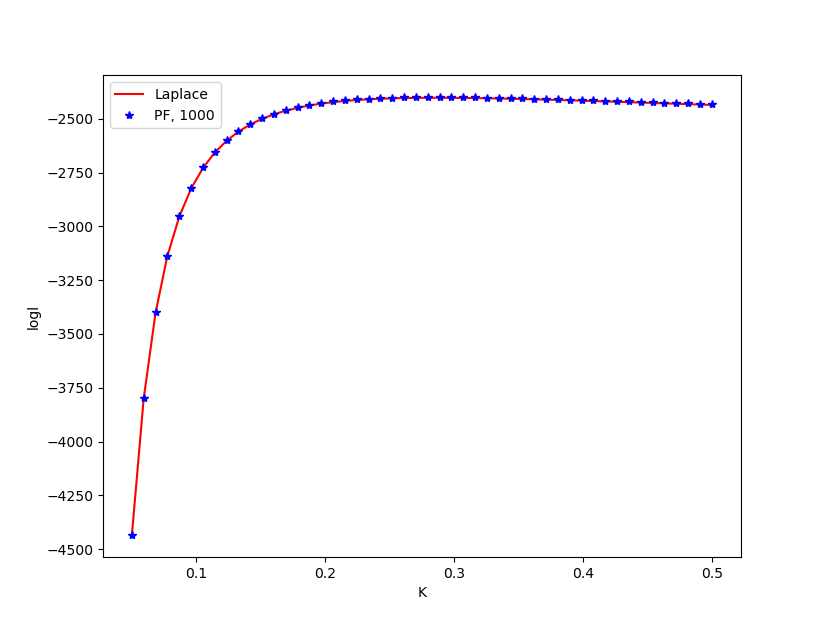}\caption{Log-likelihood profile: $K$ vs.\ log-likelihood. $A=0.7$. The log-likelihood profiles using Laplace approximation and  PF with the proposed importance sampling are shown in red line and blue stars respectively.}
 \label{fig:logl_lp_k}      
\end{figure}

\begin{figure}[ht]
 \centering
 \includegraphics[width=0.8\textwidth]{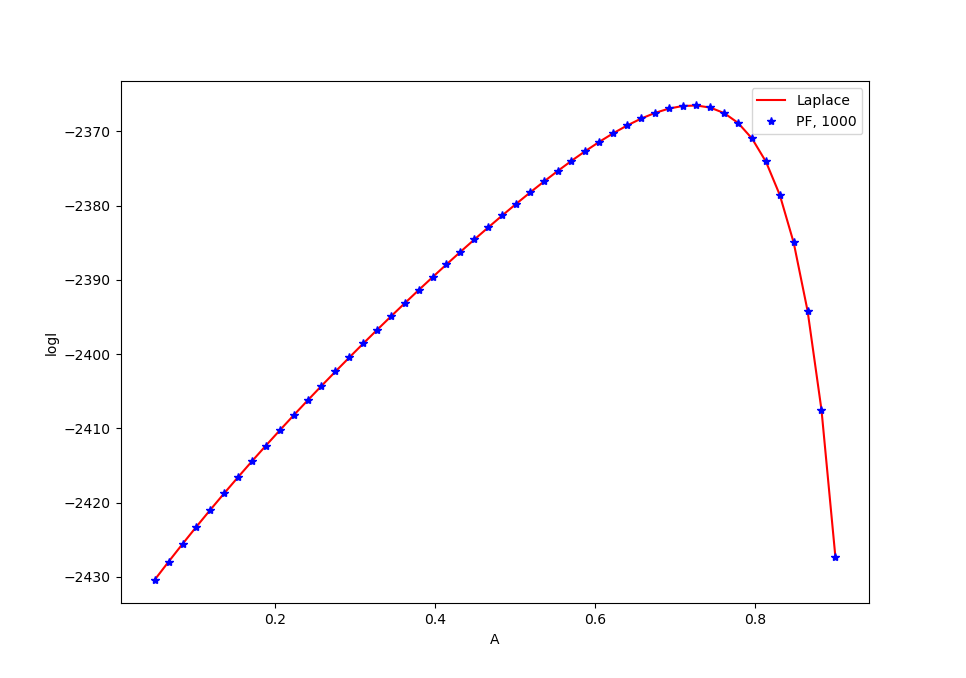}\caption{Log-likelihood profile: $A$ vs.\ log-likelihood. $K=0.3$. The log-likelihood profiles using Laplace approximation and  PF with the proposed importance sampling are shown in red line and blue stars respectively.}
 \label{fig:logl_lp_a}      
\end{figure}

The Laplace approximation is designed for high default portfolios with a relatively large number of observations.  Figures~\ref{fig:logl_lp_k} and \ref{fig:logl_lp_a} present the comparison between the Laplace approximation and the PF with importance sampling. One observes that the log-likelihood profile of the Laplace approximation matches the profile of the PF, which indicates a good precision of the Laplace approximation of the likelihood function.  

\subsection{Calibration using Laplace approximation}\label{subsection:assess2}

This section presents the calibration analysis using the Laplace approximation method. The migration model used is the two-factor transition model as in Equation~\eqref{eq:probitT}. The model parameters used for the simulation are specified as follows 
\begin{align*}
A &= \diag([a_d, a_p]) = \diag([0.7, 0.8]),\\
Q &= \begin{pmatrix}
\sqrt{1-A^2(1, 1)} &  0 \\
0 & \sqrt{1-A^2(2, 2)}
\end{pmatrix} * 
\begin{pmatrix}
1 & \rho \\
\rho & 1
\end{pmatrix} * 
\begin{pmatrix}
\sqrt{1-A^2(1, 1)} &  0 \\
0 & \sqrt{1-A^2(2, 2)}
\end{pmatrix}\\
&= \begin{pmatrix}
0.7141 &  0 \\
0 & 0.7141
\end{pmatrix} * 
\begin{pmatrix}
1 & 0.4 \\
0.4 & 1
\end{pmatrix} * 
\begin{pmatrix}
0.6 &  0 \\
0 & 0.6
\end{pmatrix}\,,
\\
K &= \diag([k_d, k_p]) = \diag([0.3, 0.2])\,.
\end{align*}
The $d$ parameter is set such that the long-term average migration probabilities are 
\begin{align*}
[\bar{T}_{1R}, \bar{T}_{2R}, \bar{T}_{3R}] &= [0.01, 0.04, 0.1]\,, \\
\bar{T}^{ND} &= 
\begin{pmatrix}
T_{11}^{ND} & T_{12}^{ND} & T_{13}^{ND}\\
T_{21}^{ND} & T_{22}^{ND} & T_{23}^{ND}\\
T_{31}^{ND} & T_{32}^{ND} & T_{33}^{ND}\\
\end{pmatrix}
=
\begin{pmatrix}
0.85 & 0.1 & 0.05 \\
0.2 & 0.6 & 0.2  \\
0.1 & 0.2 & 0.7 \\
\end{pmatrix}\,.
\end{align*}
To test the calibration method, $1000$ different scenarios of migrations are simulated based on the parameter values above. The Laplace method is applied to calibrate the model parameters $A$ and $K$ using each scenario of simulated migrations. The $d$ parameters are derived beforehand according to the average of the observed cumulative migration probabilities of each scenario, using Equation~\eqref{eq:EPhi_x}, similar to the procedure in Section~\ref{subsection:TM_model_setup}

\subsubsection{Calibration results}

Table~\ref{tab:probT} presents the statistics of the estimates in the $1000$ scenarios. One observes a very good match between the average of the estimates and the `true' parameters defined at the beginning of  Section~\ref{subsection:assess2}. The variance of the estimates mainly comes from the randomness of the Monte Carlo sampling. For example, the model assumes a unit-variance of each of the elements of the $x_k$, i.e.\ the equilibrium distribution of $x_k$ has is such that the  variance of the elements is equal to one. Hence the covariance matrix of the  idiosyncratic variable $\eta_{k}$ must then be $I-A^2$ as $A$ is diagonal. But in the simulation, the realized variance of the simulated $\eta_{k}$ could deviate from this assumption and hence impact the estimation of the $K$ parameters. To illustrate this, in the simulation, we re-normalize the Monte Carlo samples of $\eta_{k}$ so that its realized covariance matrix is always equal to $I-A^2$. The re-calibration results are presented in the Table~\ref{tab:probT_std}. One observes that the average of the $K$ estimates are closer to the `true' values and the standard deviation are much smaller. Other factors which could impact the variance of the estimates are the values of the model parameters. For example, in a one-dimensional model for the $x_k$, the bigger the autocorrelation parameter $A$ is, the bigger the autocovariance of the AR(1) process is. Consequently it creates more uncertainty in the estimation of the $K$ parameters. On the other hand, the bigger $A$ is, the smaller the variance of $\eta_k$ is (keeping the variance of the $x_k$ fixed), and hence the smaller the Monte Carlo error of the estimation of $A$ parameters is. Table~\ref{tab:probT_v2} presents the calibration results of the model with $A$ reset to $\diag([0.3, 0.4])$ and the other parameters keeping the same. Compared to Table~\ref{tab:probT}, one observes that with a smaller $A$ value the variance of the $K$ estimates decreases while the variance of the $A$ estimates increases.  

\begin{table}[ht]
\centering
\begin{tabular}{|c|c|c|c|c|c|}
\hline
 & $a_d$ & $a_p$ & $k_d$ & $k_p$ & $\rho$ \\
\hline
average & 0.6768 & 0.7732 & 0.2962 & 0.1976 & 0.3998 \\
 std    & 0.0550 & 0.0493 & 0.0264 & 0.0217 & 0.0705 \\
\hline 
\end{tabular}
\caption{Average and the standard deviation of the estimates based on the Laplace method.}
\label{tab:probT}
\end{table}

\begin{table}[ht]
\centering
\begin{tabular}{|c|c|c|c|c|c|}
\hline
 & $a_d$ & $a_p$ & $k_d$ & $k_p$ & $\rho$ \\
\hline
average & 0.6767 & 0.7732 & 0.2985 & 0.2002 & 0.3994 \\
 std    & 0.0551 & 0.0494 & 0.0086 & 0.0080 & 0.0705 \\
\hline 
\end{tabular}
\caption{Average and the standard deviation of the estimates based on the Laplace method with re-normalized $\eta_k$.}
\label{tab:probT_std}
\end{table}

\begin{table}[ht]
\centering
\begin{tabular}{|c|c|c|c|c|c|}
\hline
 & $a_d$ & $a_p$ & $k_d$ & $k_p$ & $\rho$ \\
\hline
average & 0.2887 & 0.3998 & 0.2962 & 0.1976 & 0.3998 \\
 std    & 0.0685 & 0.0703 & 0.0182 & 0.0133 & 0.0702 \\
\hline 
\end{tabular}
\caption{Average and the standard deviation of the estimates based on the Laplace method with resetting $a_d = 0.3$ and $a_p=0.4$}
\label{tab:probT_v2}
\end{table}

\subsubsection{Stepwise Calibration results}

An alternative procedure to calibrate the transition model is a stepwise approach, aimed at reducing the parameter dimensionality. Note that the conditional likelihood of the observed migrations of row $i$ of the two-factor transition model, see  Equation~\eqref{eq:probitT}, can be rewritten as, where we use $C_{i, k} = \frac{N_{i, k}!}{\prod_{j=1}^{R} m_{ij,k}!}$, 

\begin{align*}
&\log p(m_{i, k}|\theta_{k}) \\
&\quad = \sum_{j=1}^{R} m_{ij,k}\log T_{ij,k}+ \log C_{i, k} \\
&\quad= m_{iR,k}\log T_{iR,k} + \sum_{j=1}^{R-1} m_{ij,k}\log T_{ij,k} + \log C_{i, k} \\
&\quad= m_{iR,k}\log T_{iR,k} + \sum_{j=1}^{R-1} m_{ij,k}[\log(1-T_{iR,k})+\log(T_{ij,k})^{ND}] + \log C_{i, k} \\
&\quad= \underbrace{m_{iR, k}\log T_{iR, k} + (N_{i, k}-m_{iR, k})\log(1-T_{iR,k})}_{\mathrm{default\, only}} + \underbrace{\sum_{j=1}^{R-1}m_{ij, k} \log(T_{ij,k}^{ND})}_{\mathrm{migration\, given\, no \,default}} + \log C_{i, k}\,,
\end{align*}
with $N_{i, k}$ the number of clients in rating $i$ at time $k$.
Therefore, one can separate the calibration of the default and non-default migrations of the transition model. This stepwise approach works as follows.
\begin{enumerate}
\item Calibrate the default only model, with the observed number of defaults, number of clients, and PD, at time $k$ for rating $i$, equal to $m_{iR, k}$, $N_{i, k}$ and $T_{iR, k}$ respectively. The outputs of the calibration are the $a_d$ and $k_d$ parameters for the default part of the model and the latent process $x_k^{(D)}$. 
\item Calibrate the migration model, with observed number of migrations and the migration probabilities, at time $k$ for rating $i=1, \ldots, R-1$ and $j=1, \ldots, R-1$, equal to $m_{ij, k}$ and $T_{ij,k}^{ND}$ respectively. The outputss of the calibration are the $a_p$ and $k_p$ parameters for the non-default part of the model and the latent process $x_k^{(P)}$.
\item Empirically estimate the correlation parameter $\rho$ by using the calibrated parameter $A$ and the realized process $[x_k^{(D)}, x_k^{(P)}]$ obtained from step 1 and 2. 
\end{enumerate}
Tables~\ref{tab:probT_stepwise} and \ref{tab:absError} present the calibration results of the stepwise calibration and the comparison with the standard two-factor model calibration. On observes that the stepwise approach produces very similar results as the standard approach. The advantage of the stepwise approach is that it reduces the parameter dimensionality in the calibration. This can be very helpful to improve the efficiency of the calibration. 

\begin{table}[ht]
\centering
\begin{tabular}{|c|c|c|c|c|c|}
\hline
 & $a_d$ & $a_p$ & $k_d$ & $k_p$ & $\rho$ \\
\hline
average & 0.6775 & 0.7709 & 0.2901 & 0.1897 & 0.3947 \\
 std    & 0.0585 & 0.0528 & 0.0277 & 0.0226 & 0.0703 \\
\hline 
\end{tabular}
\caption{Average and the standard deviation of the estimates based on the Laplace method with the stepwise approach.}
\label{tab:probT_stepwise}
\end{table}

\begin{table}[ht]
\centering
\begin{tabular}{|c|c|c|c|c|c|}
\hline
 & $a_d$ & $a_p$ & $k_d$ & $k_p$ & $\rho$ \\
\hline
average error & 0.0058 & 0.0065 & 0.0153 & 0.0174 & 0.0049 \\
\hline 
\end{tabular}
\caption{Average absolute error of the stepwise approach compared with the standard two factor model calibration.}
\label{tab:absError}
\end{table}

\subsection{Calibration using the particle filter}\label{subsection:assess3}

\begin{figure}[ht]
 \centering
 \includegraphics[width=0.95\textwidth]{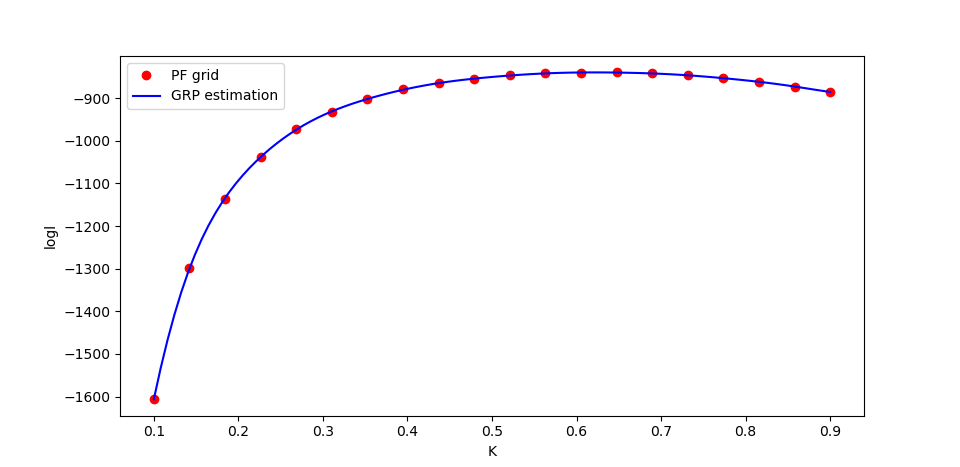}\caption{Log-likelihood profile for $k$ when $a=0.73$ The red dots are the log-likelihoods of the training grid. The blue line is the predicted log-likelihood using the trained GRP, when $a=0.73$.}
 \label{fig:GPR_fit}      
\end{figure}

In this section we test the calibration performance of the proposed particle filter (PF) with the Gaussian process regression (GPR) approach. Since the calibration is assessed using 1000 different scenarios, as in Section~\ref{subsection:assess2}, for the sake of a moderate running time, we consider the default only case, the model of \eqref{eq:probitT_DO}. But note that this calibration exercise can be rather straightforwardly extended to the cases of full migration matrix calibration. The set-up of the default only model in this experiment is the same as in Section~\ref{subsection:assess1}, namely, a high default case and a low default case. Note that although the PF calibration approach specially designed for the low default portfolio, it is in fact a generic method and can be used for both high and low default cases. In this assessment, the purpose of the calibration to a high default portfolio is to compare it with the calibration using the Laplace approach of Section~\ref{subsection:assess2}.  We build a Cartesian grid for the training GPR. The Cartesian grid is two dimensional, with one dimension for parameter $a$ and another dimension for parameter $k$. Each coordinate ranges on a set of $20$ evenly spaced points between $0.1$ and $0.9$. Therefore, in total there are $400$ pairs of points $(A, K)$ on the Cartesian grid. For each pair of points, the proposed PF Algorithm~\ref{algo:SMC_GPR} with $1000$ Monte Carlo points is run to obtain  estimates of the corresponding log-likelihood function. Then the GPR is trained on the Cartesian grid with inputs $(A, K)$ and outputs the corresponding log-likelihood. The training (and prediction) of the GPR uses the {\tt scikit-learn} Python package, with a RBF kernel (i.e., a squared-exponential kernel) plus a WhiteNoise kernel. Figure~\ref{fig:GPR_fit} shows an instance of the prediction of the trained GPR compared with the training samples. One can see a very good prediction of the log-likelihood from the trained GPR. 

\noindent Once the GPR is trained, MLE is applied to the trained GPR function to find the estimator for the model parameters $a$ and $k$. As in the experiments in Section~\ref{subsection:assess2}, $1000$ scenarios of migrations are generated based on the proposed `true' parameters. For each scenario, the PF with the GPR method is applied to calibrate the model parameters. Tables~\ref{tab:probT_smc} and \ref{tab:probT_smc_LDP} present the averages and standard deviations of the estimated parameter values based on the PF with GPR approach, for both high and low default cases. One observes that, for the high default case, the PF approach produces very similar results as the Laplace approach, which justifies the accuracy of the PF approach. It is also interesting to see that the standard deviations in the PF case are slightly higher, due to the Monte Carlo error in PF and the bias of the GPR approximation. For the low default case, the results also suggest a good performance of the PF calibration. One observes the bias in the $k$ parameter estimation and a higher standard deviation. This is mainly because of the low default nature. Namely, there is only a limited number of default cases observed and hence these introduce more estimation errors.  

\begin{table}[ht]
\centering
\begin{tabular}{|c|c|c|}
\hline
 & $a_d$  & $k_d$ \\
\hline
average & 0.6720 & 0.2903 \\
 std    & 0.0634 & 0.0290 \\
\hline 
\end{tabular}
\caption{Average and the standard deviation of the estimators from PF approach. `True' value of $k_d$ is 0.3.}
\label{tab:probT_smc}
\end{table}

\begin{table}[ht]
\centering
\begin{tabular}{|c|c|c|}
\hline
 & $a_d$  & $k_d$ \\
\hline
average & 0.7211 & 0.5518 \\
 std    & 0.0714 & 0.0993 \\
\hline 
\end{tabular}
\caption{Average and the standard deviation of the estimators of the PF approach for the Low-default case. `True' value of $k_d$ is 0.6.}
\label{tab:probT_smc_LDP}
\end{table}

\section{Conclusions}\label{section:conclusions}

This paper proposes two calibration algorithms, tailor-made for the credit rating transition models of high- and low-default portfolios respectively. The algorithm for the high-default portfolio uses the Laplace method, incorporated with the Kalman filter algorithm, which is used to estimate the mode of the posterior distribution of the signals 
and compute the (approximated) likelihood function of the observed transition matrices. We show that the same Kalman filter algorithm can be applied to both mode estimation and likelihood calculation. Therefore the numerical ML algorithm is very fast. For the low-default portfolio, since the number of the observed migrations (defaults) is limited, the performance of the Laplace approximation is not assured and hence might introduce biased in the likelihood approximations and ML estimates. As an alternative, we develop an algorithm based on a particle filter with importance sampling. The importance density is obtained from the mode estimation in the Laplace algorithm. Experiments show that mode estimation produces very accurate estimates of the posterior of the latent states and as a result, the designed importance sampling significantly improves the efficiency of the particle filter. Moreover, the GPR is used to smooth the likelihood function approximated by the particle filter, so that a numerical optimization algorithm can be easily applied to obtain the ML estimates. The efficiency and the accuracy of these two proposed calibration algorithms are substantiated by numerical experiments.

\appendix

\section{State space model}\label{appendix:ssm}
State-space models deal with dynamic time series problems that involve unobserved variables or parameters that describe the evolution of the state of the underlying system. The general {\it state space} model, defined on some probability space $\pspace$, is as follows
\begin{equation}
\begin{aligned}
x_k &=f_k(x_{k-1},u_k),\,x_0 \\
y_k  &=h_k(x_k,v_k)\,,  \quad k \in \mathbb{N}^+\,,
\end{aligned}
\label{state_space_model}
\end{equation}
where
$f_k:\Real^d \times \Real^p \rightarrow \Real^d$, $h_k: \Real^d \times \Real^q \rightarrow \Real^m$ are given Borel measurable functions, $\{u_k\}_{k\in \mathbb{N}^+}$ is a $p$-dimensional
and $\{v_k\}_{k\in \mathbb{N}^+}$ is a $q$-dimensional white noise processes both independent of the initial condition $x_0$, and mutually independent as well. Parameters in the functions $f_k$ and $h_k$, together with the covariance of $u_k$ and $v_k$ can be seen as the parameters of the state space model, to which we collectively refer to as $\psi$.

\section{Bayesian filter}\label{appendix:BF}

The Baysian filters, see for instance \citet{Press} and \citet{Robert}, are often used to estimate the latent process $(x_k)_{k=1, \ldots, n}$  or to compute the likelihood function of the observations  $y_k$ in a state space model~\eqref{state_space_model} given the model parameters. We define the initial density function $p_{0}$ of $x_{0}$. The methodology in Bayesian filtering consists of two parts: prediction and update. At every time point $k$, the prediction part computes (estimates) the {\it prior distribution (density)} of $x_k$ (a time $k$ given the past observations up to time $k-1$), 
\begin{equation*}
p(x_k\mid y_{1:k-1},\psi)
\end{equation*}
and the update part computes (estimates) the {\it posterior distribution (density)} of $x_k$ given the past up to time $k$, 
\begin{equation*}
p(x_k\mid y_{1:k}, \psi)\,.
\end{equation*}
\noindent It follows that the state space model~\eqref{state_space_model} satisfies the properties of a stochastic system, i.e.\ at every (present) time $k\geq 1$ the future states  and future observations $(x_j,y_j)$, $j\geq k$, are conditionally independent from the past states and observations $(x_j,y_{j-1})$, $j\leq k$,  given the present state $x_k$, see~\citet{vanschuppen1989}.
It then follows that  $(x_k)_{k \in \mathbb{N}}$ is a Markov process,
and for every $k\geq 1$ one has that $y_k$ and $y_{1:k-1}$ are conditionally independent given $x_{k-1}$, in terms of densities,
\begin{equation}\label{eq:xyk}
p(x_k\mid x_{k-1},y_{1:k-1}, \psi)=p(x_k\mid x_{k-1}, \psi).
\end{equation}
Similarly, due to the Markov property, $x_k$ and $y_{k+1:T}$ are independent given $x_{k+1}$, which gives
\begin{equation}\label{eq:xyk+1}
p(x_k\mid x_{k+1},y_{1:T}, \psi)=p(x_k\mid x_{k+1}, y_{1:k}, \psi).
\end{equation}
Moreover, one also has, for every $k\geq 1$,  that $y_k$ and $y_{1:k-1}$ are conditionally independent given $x_{k-1}$, in terms of densities,
 \begin{align}\label{eq:cond-ind-y}
 p(y_k\mid y_{1:k-1},x_k, \psi)=p(y_k\mid x_k, \psi)\,,\quad  \mbox{for}\quad  k \in \mathbb{N}^+\,.
\end{align}
These three equations ~\eqref{eq:xyk}, ~\eqref{eq:xyk+1} and ~\eqref{eq:cond-ind-y} are used later in the derivation of the Bayesian filter and smoother recursions. 

Using Bayes' rule and equation~\eqref{eq:xyk}, we deduce that the density function of the prior distribution is given by
\begin{equation}\label{eq:prior}
\begin{aligned}
p(x_k\mid y_{1:k-1}, \psi) &= \int p(x_k\mid x_{k-1},y_{1:k-1}, \psi)p(x_{k-1}\mid y_{1:k-1}, \psi)\, \ud x_{k-1}\\
                           &= \int p(x_k\mid x_{k-1}, \psi)p(x_{k-1}\mid y_{1:k-1}, \psi)\, \ud x_{k-1}\,.
\end{aligned}
\end{equation}
Note that when $k=1$, the posterior distribution $p(x_{k-1}\mid y_{1:k-1}, \psi)$ is defined as the initial density $p_{0}$. 

The purpose of the Bayesian algorithm is to sequentially compute the posterior distribution  $p(x_k\mid y_{1:k}, \psi)$. Again using Bayes' rule, \eqref{eq:cond-ind-y} and \eqref{eq:prior}, we obtain the posterior 
\begin{equation}\label{eq:posterior}
\begin{aligned}
p(x_k\mid y_{1:k}, \psi) &= \frac{p(y_k\mid x_k,y_{1:k-1}, \psi)p(x_k\mid y_{1:k-1}, \psi)}{p(y_k\mid y_{1:k-1}, \psi)}\\
& = \frac{p(y_k\mid x_k, \psi)p(x_k\mid y_{1:k-1}, \psi)}{p(y_k\mid y_{1:k-1}, \psi)}.\\
\end{aligned}
\end{equation}
with the conditional likelihood 
\begin{equation}\label{eq:cond_l}
p(y_k\mid y_{1:k-1}, \psi)= \int p(y_k\mid x_k, \psi)p(x_k\mid y_{1:k-1}, \psi)\, \ud x_k\,.
\end{equation}
\noindent If we assume the marginal likelihood function $p(y_k\mid x_k, \psi)$ and the transition probability $ p(x_k\mid x_{k-1}, \psi)$ are known, then given the posterior distribution $p(x_{k-1}\mid y_{1:k-1}, \psi)$ at time $k-1$ , we can use equations \eqref{eq:prior} and \eqref{eq:posterior} to compute the posterior measure $p(x_{k}\mid y_{1:k}, \psi)$ at time $k$. In this way the posterior distributions can be computed recursively given the initial distribution $p_0$.
\medskip\\
While the mathematical derivations of the Bayesian filter and smoother are straight forward, in practice it is always a big challenge to compute the integrals in the equations \eqref{eq:prior}
or \eqref{eq:cond_l}. Although there are some special cases where theoretical formulas are available for the integrals, such as the Kalman filter or (more general) with conjugate priors, in most cases one has to find approximations or numerical algorithms to compute these integrals. In the next two subsections, we introduce two types of the most used Bayesian filters in practice, the Kalman filter and its extensions, and the particle filter.

\subsection{Kalman filter}\label{subsection:KF}
Assume that the state and observations in \eqref{state_space_model} evolve according to a linear Gaussian model. That is the functions $f_k$ and $h_k$, for $k=1,\ldots, n,$, have to take linear forms as follows
\begin{equation}
\begin{aligned}
y_k & = K_k x_k + \varepsilon_k,  \\
x_{k+1} & = A_k x_{k} + \eta_{k},
\end{aligned}
\label{eq:SSM_LG}
\end{equation}
where $K_k$ is a $m \times d$ matrix and  $A_k$ is a $d \times d$ matrix and the following distributional assumptions are made. The initial variable $x_0$ is assumed to be Gaussian, the 
$\varepsilon_k$ and $\eta_k$ are assumed to be serially independent and independent of each
other at all time points, with Gaussian distributions, for $k=1,\ldots, n,$
\begin{align*}
\varepsilon_k & \sim N(0,H_k), \\
\eta_k & \sim N(0,Q_k)\,, \\
x_0 & \sim N(a_0, P_0), 
\end{align*}
in which $N(\mu, \Sigma)$ denotes the Gaussian distribution with mean $\mu$ and covariance $\Sigma$. We also use the generic notation $N(x; \mu, \Sigma)$ to denote the density at $x$ of this normal distribution. Due to the Gaussian assumptions and the linear structure of the model in \eqref{eq:SSM_LG}, the prior and posterior distribution defined in \eqref{eq:prior} and \eqref{eq:posterior} are Gaussian and can be analytically derived. Denote the estimates for the prior and posterior density at time $k$ by $N(x; x_{k|k-1}, P_{k|k-1})$ and $N(x; x_{k|k}, P_{k|k})$  respectively. Here is the Kalman filter algorithm for the Gaussian linear model~\eqref{eq:SSM_LG}.

\begin{algorithm}[{\bf Kalman filter}]\label{algo:KF}
\mbox{}\vspace{-0.0em}
\begin{description}
\item[Initialization:] Initialize the posterior distribution $p_0(x_0)= N(x_0; x_{0|0}, P_{0|0}).$ 
\item[Iteration:] For $k=1, \ldots, n$, let $x_{k-1|k-1}$ and $P_{k-1|k-1}$ be given.
\begin{enumerate}
\item Compute the prior distribution $p(x_k|y_{1:k-1})=N(x_k; x_{k|k-1}, P_{k|k-1})$, with
\begin{equation}
\begin{aligned}
x_{k|k-1} &= A_k x_{k-1|k-1},  \\
P_{k|k-1} & = A_kP_{k-1|k-1}A_k^\top + Q_k, \end{aligned}
\end{equation}
\item With $S_k= K_kP_{k|k-1}K_k^\top + H_k$, compute the Kalman gain 
\begin{equation}
G_k = P_{k|k-1}K_k^\top(S_k)^{-1}\,. 
\end{equation}
\item Compute the posterior distribution $p(x_k|y_{1:k})=N(x_k; x_{k|k}, P_{k|k})$, with
\begin{equation}\label{eq:posterior2}
\begin{aligned}
x_{k|k} &= x_{k|k-1} + G_k(y_k-K_kx_{k|k-1}), \\
P_{k|k} &= P_{k|k-1} - G_kK_kP_{k|k-1}. 
\end{aligned}
\end{equation}

\item Compute the the conditional log-likelihood  $p(y_{k}|y_{1:k-1})$ as
\begin{equation*}
p(y_{k}|y_{1:k-1}) = N(y_k; H_kx_{k|k-1}, S_k).  
\end{equation*}
\end{enumerate}
\item [Likelihood:] The log likelihood of the observations, $\log p(y_{1:n})$, can be obtained by summing up the conditional log-likelihoods, 
\begin{equation*}
\log p(y_{1:n}) = \sum_{k=1}^n\log p(y_{k}|y_{1:k-1}) \,. 
\end{equation*}

\end{description}
\end{algorithm}

\noindent Extensions and generalizations of the Kalman filter approach have also been developed, such as the extended Kalman filter, see~\citet{einicke1999robust} or \citet{wan2001dual}, and the unscented Kalman filter, see~\citet{wan2001unscented}, which work for nonlinear systems. However the extended Kalman filter and the unscented Kalman filter do not account for the non-Gaussian properties of the model other than their first two moments, the mean and variance functions. \citet[Section~10.6-10.7]{durbin2012time} proposes a mode estimation approach which captures non-Gaussian properties of the state space model. 

  
\subsection{Particle filter}\label{subsection:PF}
In the {\it particle filter}, the prior and posterior distributions are estimated by a Monte Carlo method. With a Monte Carlo method, a certain density function $f(x)$ is generally estimated by 
\begin{equation*}
f(x)= \sum_{i=1}^{N}w^{(i)}\delta_{x^{(i)}}(x)\,,
\end{equation*}
where $\{x^{(i)}, i=1,\ldots,N\}$ are i.i.d.~random samples from a so-called {\it importance density}, $\{w^{(i)}, i=1,\ldots,N\}$ are {\it the importance weights}, and $\delta_{x^{(i)}}$ are Dirac distributions. The key parts of the particle filter is to choose the importance density and to compute the importance weights, see e.g.~\citet{D}. For the general state space model (\ref{state_space_model}), suppose the posterior density $p(x_{k-1}|y_{1:k-1})$ at time ${k-1}$ is estimated by
\begin{equation*}
p(x_{k-1}|y_{1:k-1}, \psi)\approx \sum_{i=1}^{N}w_{k-1}^{(i)}\delta_{x_{k-1}^{(i)}(x_{k-1})}.
\end{equation*}
Using Equations~\eqref{eq:prior} and~ \eqref{eq:posterior},
the prior and posterior densities are respectively estimated by 
\begin{equation}\label{eq:PF_nosampling}
\begin{aligned}
p(x_{k}|y_{1:k-1}, \psi) &\approx \sum_{i=1}^N w_{k-1}^{(i)}p(x_k|x_{k-1}^{(i)}, \psi),\\
p(x_{k}|y_{1:k}, \psi) &\approx \frac{\sum_{i=1}^N w_{k-1}^{(i)}p(y_k|x_k, \psi)p(x_k|x_{k-1}^{(i)}, \psi)}{\int \sum_{i=1}^N w_{k-1}^{(i)}p(y_k|x_k, \psi)p(x_k|x_{k-1}^{(i)}, \psi) \ud x_{k} }.
\end{aligned}
\end{equation}
These estimated densities are mixture distributions. If the samples $\tilde{x}_k^{(i)}$ are generated from the transition density $p(x_k\mid x_{k-1}^{(i)}, \psi)$ for $i=1,\ldots,N$, the estimates of prior and posterior in Equation~\eqref{eq:PF_nosampling} can be reformulated so that they can be used in recursive calculations,
\begin{equation*}
\begin{aligned}
p(x_{k}|y_{1:k-1}, \psi) &\approx \sum_{i=1}^N w_{k-1}^{(i)}\delta_{\tilde{x}_k^{(i)}}(x_k),\\
p(x_{k}|y_{1:k}, \psi) &\approx \sum_{i=1}^N w_{k}^{(i)}\delta_{\tilde{x}_k^{(i)}}(x_k)\,,
\end{aligned}
\end{equation*}
where the weights $w_{k}^{(i)}$ are defined by
\begin{equation}\label{eq:w_PF}
w_{k}^{(i)} = \frac{w_{k-1}^{(i)}p(y_k\mid \tilde{x}_k^{(i)}, \psi)}{\sum_{i=1}^N w_{k-1}^{(i)}p(y_k\mid \tilde{x}_k^{(i)}, \psi)}. 
\end{equation}
Consequently one obtains the conditional likelihood $p(y_{k}|y_{1:k-1})\approx \sum_{i=1}^N w_{k-1}^{(i)} p(y_k|\tilde{x}_k^{(i)}, \psi)$.

This type of particle filter is often referred to as sequential Monte Carlo. It is a specific member of the family termed the bootstrap particle filter, see \citet{gordon1993novel}. In \citet{D} it is shown that the variance of the importance weights decreases
stochastically over time. This will lead the importance weights to be concentrated on a
small amount of sampled particles. This problem is called degeneracy. To address the rapid degeneracy problem, the sampling-importance resampling (SIR) method, see e.g., \citet{D} and \citet{PS}, is introduced to eliminate the samples with a low importance weight and multiply the samples with a high importance weight. In SIR, once the approximation
of the posterior $p(x_{k}|y_{1:k}, \psi) \approx \sum_{i=1}^N w_{k}^{(i)}\delta_{\tilde{x}_k^{(i)}}(x_k)$ is obtained, new, re-sampled, particles $x_k^{(j)}$ are
i.i.d.\ sampled from this approximate posterior distribution, i.e., every $x_k^{(j)}$ is independently chosen from
the $\tilde{x}_k^{(i)}$ with probabilities $w_k^{(i)}$ for $i = 1,\ldots, N$. This step can be accomplished by sampling integers $j$ from $\{1,\ldots, n\}$ with probabilities $w_k^{(i)}$ for $i = 1,\ldots, N$. Then the new estimation on the posterior and conditional likelihood is given by 
\begin{equation*}
\begin{aligned}
p(x_{k}|y_{1:k}, \psi) &\approx \frac{1}{N}\sum_{i=1}^N \delta_{\tilde{x}_k^{(j)}}(x_k)\,, \\
p(y_{k}|y_{1:k-1}, \psi) &\approx \frac{1}{N} \sum_{i=1}^N  p(y_k|x_k^{(j)})\,.
\end{aligned}
\end{equation*}

\section{Gaussian process regression (GPR)}\label{appendix:GPR}
In this section, we provide a brief introduction to {\it Gaussian process regression} (GPR). For a good overview, we refer for example to \citet{RW}.
We start by introducing Gaussian processes.
\begin{defn}\label{def:GP}
For any index set $\chi$, a {\it Gaussian process} (GP) on $\chi$ is a set of random variables $(f(x), x\in \chi)$ such that $\forall n \in \mathbb{N}$ and $x_1,\ldots,x_n \in \chi$, $(f(x_1),\ldots, f(x_n))$ is a 
multivariate Gaussian random variable.
\end{defn}
\noindent
In Definition~\ref{def:GP}, $f:\chi \times \Omega \rightarrow \mathbb{R}$ and the index set $\chi$
is the set of possible inputs. From now on, we specify $\chi$ to be $\mathbb{R}^d$.
\medskip\\
A GP is completely specified by its {\it mean} and {\it covariance} functions. We define the mean function $m:\mathbb{R}^d \rightarrow \mathbb{R}$ and the covariance $k:\mathbb{R}^d \times \mathbb{R}^d \rightarrow\mathbb{R}$ of $f$ as 
\begin{align*}\label{eq:mean-kernel}
m({\bf x)} &=\mathbb{E}[f({\bf x})]\,,\nonumber\\
k({\bf x},{\bf x}') &= \mathbb{E}[(f({\bf x}) -m({\bf x})][f({\bf x}') -m({\bf x}')]\,.
\end{align*}
\noindent Given $n$ pairs of observations $({\bf x}_1, y_1), \ldots, ({\bf x}_n, y_n)$, ${\bf x}_i \in \mathbb{R}^d$ and $y_i \in \mathbb{R}$ for $i=1,\ldots,n$, the GPR model is a probabilistic non-parametric model with the following structure
\begin{equation*}
Y = \mathbf{f}(X) + \varepsilon\,,
\end{equation*}
where $Y= [y_1,y_2,\ldots,y_n]^T$ are the outputs, $X= [{\bf x}_1, {\bf x}_2,\ldots, {\bf x}_n]^T$ are the inputs, $\varepsilon$ is a Gaussian noise with a vector mean ${\bf 0}$ and variance $\sigma^2 \mathbf{I}$, and $\mathbf{f}: \mathbb{R}^d\times \cdots \times \mathbb{R}^d\rightarrow \mathbb{R}^n$ is such that $\mathbf{f}(X) = [f({\bf x}_1), f({\bf x}_2), \ldots,f({\bf x}_n)]^T$ follows a multivariate Gaussian distribution specified by the GP mean $m(x)$ and covariance functions $k({\bf x}_i,{\bf x}_j)$, $1\leq i,j\leq n$. Namely $\mathbf{f}(X) \sim N({\pmb \mu}, K)$, where $K$ is the $n\times n$ covariance matrix of which the $(i, j)$-th element $K_{ij} = k({\bf x}_i, {\bf x}_j)$ and for simplicity, we take the mean vector ${\pmb \mu} ={\bf 0}$. 

To predict $Y_* = [y_{*1}, \ldots,y_{*m}]^T$ at the test locations $X_* = [{\bf x}_{n+1}, \ldots,{\bf x}_{n+m}]^T$, the joint distribution of the training observations $Y$ and the 
predictive targets $Y_*$ are given by
\begin{equation}
\label{GPR_joint}
\left[\begin{array}{c}
  Y \\
  Y_*
\end{array}\right] \sim
N\left(\mathbf{0}, \left[\begin{array}{cc}
                  K(X,X)+ \sigma^2 \mathbf{I} & K(X,X_*) \\
                  K(X_*,X) & K(X_*,X_*) \\
                \end{array}
                 \right]\right)\,,
\end{equation}
where $K(X, X) = K$, $K(X_*, X)$ is an $m \times n$ matrix with $(i, j)$-th element $[K(X_*, X)]_{ij} = k({\bf x}_{n+i}, {\bf x}_j)$. The other entries $K(X,X_*)$ and $K(X_*, X_*)$ are defined analogously.
Hence the predictive distribution is also Gaussian. Namely,
\begin{equation*}
\label{GPR_condtional}
Y_*\mid Y,X,X_* \sim  N\left(\tilde{Y}, P \right)\,,
\end{equation*}
with
\begin{align}
\tilde{Y} &= K(X_*,X)[K(X,X)+\sigma^2{\bf I}]^{-1}Y\,,\label{GPR_mean}\\
P&=K(X_*,X_*)-K(X_*,X)[K(X,X)+\sigma^2{\bf I}]^{-1}K(X,X_*)\,.\label{GPR_variance}
\end{align}
In view of \eqref{GPR_mean} and \eqref{GPR_variance}, the {\it covariance/kernel} function $k$ plays a significant role in GPR. One commonly used kernel is the {\it squared exponential}, i.e.,
\begin{equation}\label{squared-exponential}
k({\bf x}_i,{\bf x}_j) = \sigma_f^2 e^{-\frac{1}{2}({\bf x}_i-{\bf x}_j)^\top\Lambda({\bf x}_i-{\bf x}_j)}\,, \qquad i,i=1,\cdots,n\,,
\end{equation}
where $\sigma_f \in \mathbb{R}$ and $\Lambda$ is a $d\times d$ positive-definite matrix. For more examples of Kernel functions and an in-depth analysis on 
how choosing this function, we refer to \citet[Chapter~4]{RW}. Each kernel has a number of parameters which specify the precise
shape of the covariance function. These are referred to as {\it hyperparameters},
since they can be viewed as specifying a distribution over function parameters, instead of
being parameters which specify a function directly. The hyperparameters in the squared exponential kernel are $\sigma_f$ and the entries of the matrix $\Lambda$.
\subsection{Computation of the predictive targets}

\noindent Denote the vector $[K(X,X)+\sigma^2{\bf I}]^{-1}Y$ in \eqref{GPR_mean} by $\alpha = [\alpha_1,\cdots,\alpha_n]^T$, then we have the estimation for $Y_*$
\begin{align}
\label{eq: GPR_estimation_linear}
Y_{*j} \approx\tilde{Y}_j &= \sum_{i=1}^n\alpha_i k({\bf x}_{n +i},{\bf x}_j)\,, \qquad j=1,\cdots, n\,.
\end{align}
If the Gaussian kernel is known, the value $Y_*$ at a point $X_*$ can be estimated by \eqref{eq: GPR_estimation_linear}. 

\subsection{Training the GPR}

Training the GPR means to determine the hyperparameters using the data $(X,Y)$. The training is usually conducted by using maximum likelihood estimation.  
Observing that $Y \sim N({\bf 0}, K+\sigma^2 {\bf I})$, we deduce that the log likelihood is given by
\begin{equation}\label{eq:loglikelihood-Y}
\log p(Y|X) = -\frac{1}{2} Y^T(K+\sigma^2{\bf I})^{-1} Y -\frac{1}{2}\log|K + \sigma^2 {\bf I}| -\frac{n}{2} \log 2\pi\,.
\end{equation}
\subsubsection{Train the GPR on Cartesian grid}
The computational complexity for computing the marginal likelihood in equation \eqref{eq:loglikelihood-Y}
is dominated by the need to invert the $K$ matrix. Standard methods for matrix
inversion of positive definite symmetric matrices require a time of order $O(n^3)$ for
inversion of an $n$ by $n$ matrix. If the valuation set is very big, the computational time could be significant. When the covariance function is expressible as a tensor product kernel and the inputs form a multidimensional grid, it was shown in \citet{saatcci2012scalable} that the costs for the GPR can be reduced (see also \citet{xu2011infinite}, \citet{luo2013fast}). We briefly describe in the sequel the GPR on multidimensional grids.
\medskip\\
We assume the kernel $k: \mathbb{R}^d\times \mathbb{R}^d \rightarrow \mathbb{R}$ is written in terms of kernels $k_m: \mathbb{R}\times \mathbb{R} \rightarrow \mathbb{R}$, for $m=1,\ldots,d$ as follows
\begin{align}\label{Kernel_ind}
k({\bf x}_i, {\bf x}_j) = \Pi_{m=1}^ d k_m( x_i^{(m)},  x_j^{(m)})\,.
\end{align}
Consider the squared exponential kernel as in \eqref{squared-exponential} with a 
diagonal covariance $\Lambda$, i.e., $\Lambda = \diag(\frac{1}{\ell_1^2},\cdots,\frac{1}{\ell_d^2})$, for $\ell \in \mathbb{R}$. It holds
\begin{align*}
k({\bf x}_i, {\bf x}_j) &= \sigma_f^2 \exp{\left\{-\sum_{m=1}^d \frac{(x_i^{(m)}-x_j^{(m)})^2}{2\ell_k^2}\right\}}\\
&= \Pi_{m=1}^d k_m(x_i^{(m)},x_j^{(m)})\,,
\end{align*}
where $k_m(x,y) = \sigma_f^{2/d} \exp{\{-(x-y)^2/2\ell_m^2\}}$.

We consider the data on a $d$-dimensional Cartesian grid. Let $\{n_s, s=1,\ldots,d)$ be the number of points in each dimension in the grid. Then the grid has $n=n_1\times \cdots\times n_d$ points. Let $X$ be the list of $n$ points in the grid, i.e., $X = ({\bf x}_1,\cdots, {\bf x}_n)$ and the indexes $n_1, \ldots, n_d$ are such that $1\leq i_1\leq n_1$, $1\leq i_2\leq n_2$, $\cdots$, $1\leq i_d\leq n_d$. The points ${\bf x}_j$, $1\leq j \leq n$, have the form ${\bf x}_j = (x_{j(1)}^{(1)},x_{j(2)}^{(2)},\cdots x_{j(d)}^{(d)})$, with $x_{j(s)}^{(s)}$ being the $j(s)$-th element in the dimension $s$ in the grid.
 We put the points of the Cartesian grid in the following order 
\begin{equation}
\label{Cartesian}
{\bf x}_{(i_1-1)\times n_2\times \cdots \times n_d + \cdots + (i_d-1)\times n_d+ i_d} = (x_{i_1}^{(1)},\ldots,x_{i_d}^{(d)}).
\end{equation}
Consider for example, a three-dimensional Cartesian grid. Take $n_1=3$, $n_2=n_3=2$. Then the grid has $12$ points and equation \eqref{Cartesian} yields
\begin{align*}
{\bf x}_1 &= (x_1^{(1)}, x_1^{(2)}, x_1^{(3)})\,,\\
{\bf x}_2 &= (x_1^{(1)}, x_1^{(2)}, x_2^{(3)})\,,\\
{\bf x}_3 &=(x_1^{(1)}, x_2^{(2)}, x_1^{(3)})\,,\\
\vdots\\
{\bf x}_{12} &=(x_3^{(1)}, x_2^{(2)}, x_2^{(3)})\,.
\end{align*}
For the GPR model, the Gaussian kernel on the Cartesian grid is given by $K_{i,j} = k({\bf x}_i, {\bf x_j})$. Suppose the kernel function $k$ has property (\ref{Kernel_ind}).
Then \eqref{Cartesian} becomes
\begin{equation*}
\begin{aligned}
&K_{(i_1-1)\times n_2\times \cdots \times n_d + \cdots +  i_d, (j_1-1)\times n_2\times \cdots \times n_d + \cdots + j_d }\\
 &\qquad = k([x_{i_1}^{(1)},\ldots,x_{i_d}^{(d)}],[x_{j_1}^{(1)},\ldots,x_{j_d}^{(d)}])\\
&\qquad= \Pi_{k=1}^d k_k(x_{i_k}^{(k)},x_{j_k}^{(k)})\\
&\qquad= k_1(x_{i_1}^{(1)},x_{j_1}^{(1)})\times \cdots \times k_d(x_{i_d}^{(d)}, x_{j_d}^{(d)})\,.
\end{aligned}
\end{equation*}
Let the matrix Kernel  $K_k = (k_k(x_i^{(k)},x_j^{(k)}))_{1\leq i,j\leq n_k}$ and recall $K$ in \eqref{GPR_joint}. We obtain,
\begin{equation*}
K = K(X, X) =  \otimes_{k=1}^d K_k\,,
\end{equation*}
where $\otimes$ denotes the Kronecker product. It follows that
\begin{equation*}
K^{-1} = \otimes_{k=1}^d K_k^{-1}\,.
\end{equation*}
So instead of inverting a high dimensional matrix, one needs to invert $d$ smaller matrices, and this reduces the computational time significantly.

\bibliographystyle{plainnat} 
\bibliography{transition_matix_calibration}

\end{document}